\documentclass[%
reprint, 
amsmath,
amssymb,
aps,
pra,
floatfix,
]{revtex4-1}

\usepackage{graphicx}
\usepackage{dcolumn}
\usepackage{bm}
\usepackage{physics}
\usepackage[usenames, dvipsnames]{color}

\begin{document}
\title{One-second coherence for a single electron spin coupled to \\ a multi-qubit nuclear-spin environment}

\author{M. H. Abobeih$^{1,2}$}
\author{J. Cramer$^{1,2}$}
\author{M. A. Bakker$^{1,2}$} 
 \thanks{Current address:  Department of Electrical Engineering and Computer Science, Massachusetts Institute of Technology, Cambridge, Massachusetts, United States.}
\author{N. Kalb$^{1,2}$}
\author{M. Markham$^3$} 
\author{D. J. Twitchen$^3$}
\author{T. H. Taminiau$^{1,2}$}
 \email{T.H.Taminiau@TUDelft.nl}

\affiliation{$^{1}$QuTech, Delft University of Technology, PO Box 5046, 2600 GA Delft, The Netherlands}
\affiliation{$^{2}$Kavli Institute of Nanoscience Delft, Delft University of Technology, PO Box 5046, 2600 GA Delft, The Netherlands}
\affiliation{$^{3}$Element Six Innovation, Fermi Avenue, Harwell Oxford, Didcot, Oxfordshire OX11 0QR, United Kingdom}

\date{\today}

\begin{abstract} 
Single electron spins coupled to multiple nuclear spins provide promising multi-qubit registers for quantum sensing and quantum networks. The obtainable level of control is determined by how well the electron spin can be selectively coupled to, and decoupled from, the surrounding nuclear spins. Here we realize a coherence time exceeding a second for a single electron spin through decoupling sequences tailored to its microscopic nuclear-spin environment. We first use the electron spin to probe the environment, which is accurately described by seven individual and six pairs of coupled carbon-13 spins. We develop initialization, control and readout of the carbon-13 pairs in order to directly reveal their atomic structure. We then exploit this knowledge to store quantum states for over a second by carefully avoiding unwanted interactions. These results provide a proof-of-principle for quantum sensing of complex multi-spin systems and an opportunity for multi-qubit quantum registers with long coherence times.
\end{abstract}
\maketitle

\begin{figure}[tb]
\includegraphics[scale=0.75]{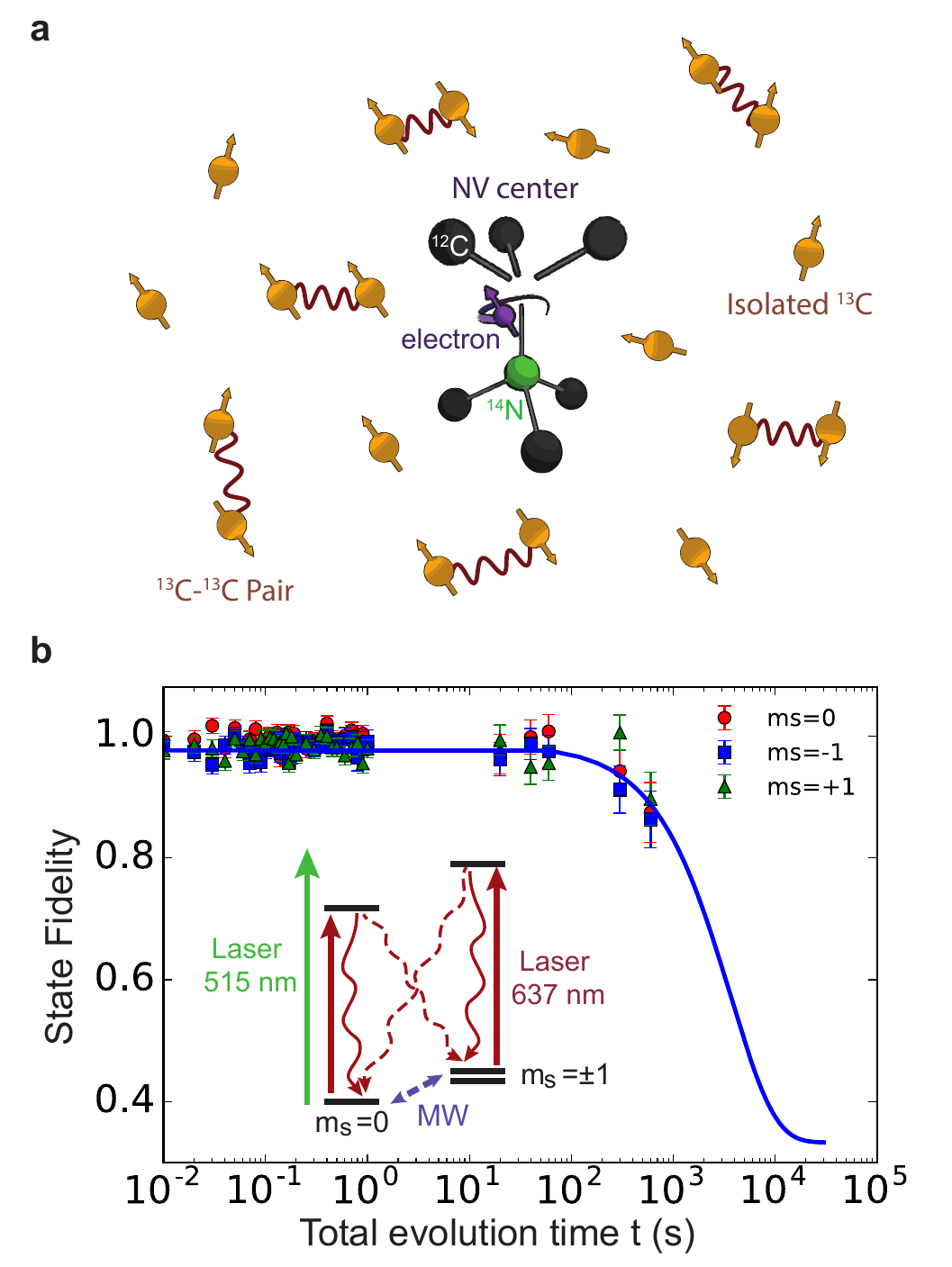}
\caption{\label{Figure1}
\textbf{Experimental system and $T_1$ measurements.} (a) We study a single nitrogen-vacancy (NV) center in diamond surrounded by a bath of $^{13}C$ nuclear spins ($1.1\%$ abundance). In this work, we show that the microscopic nuclear-spin environment is accurately described by $7$ isolated $^{13}C$ spins, $6$ pairs of coupled $^{13}C$ spins and a background bath of $^{13}C$ spins (not depicted). (b) Longitudinal relaxation of the NV electron spin. The spin is prepared in $m_s = 0, -1$, or $+1$ and the fidelity with the initial state is measured after time $t$. The inset shows the microwave (MW) and laser controls for the NV spin and charge states, as well as the pathways for spin relaxation induced by potential background noise from these controls.}
    \end{figure}
    
Coupled systems of individual electron and nuclear spins in solids are a promising platform for quantum information processing \cite{Pfaff_NatPhys2013, Waldherr_Nature2014, Cramer_NatureComm2016, Wolfowicz_NJP2016,Muhonen_NatNano2014,dehollain2016bell} and quantum sensing \cite{Zaier_NatComm2016,Pfender_arXiv2017,Rosskopf_NPJQI2017,Lovchinsky_Science2016, Unden_PRL2016}. Initial experiments have demonstrated the detection and control of several nuclear spins surrounding individual defect or donor electron spins \cite{Kolkowitz_PRL2012, Taminiau_PRL2012, Zhao_NatureNano2012, Muller_NatComm2014, Shi_NatPhys2014,Lee_NatNano2013}. These nuclear spins provide robust qubits that enable enhanced quantum sensing protocols \cite{Zaier_NatComm2016,Pfender_arXiv2017,Rosskopf_NPJQI2017,Lovchinsky_Science2016, Unden_PRL2016}, quantum error correction \cite{Waldherr_Nature2014, Cramer_NatureComm2016, Kalb_NatureComm2016}, and multi-qubit nodes for optically connected quantum networks \cite{Hensen_Nature2015, Sen_NatPhoton2016, Reiserer_PRX2016, Kalb_Science2017}. 

The level of control that can be obtained is determined by the electron spin coherence and therefore by how well the electron can be decoupled from unwanted interactions with its spin environment. Electron coherence times up to $0.56$ s for a single electron spin qubit \cite{Muhonen_NatNano2014} and $\sim 3$ seconds for ensembles \cite{Bar-Gill_NatComm2012, Tyryshkin_NatMat2012, Wolfowicz_PRB2012, Wolfowicz_NatNano2013} have been demonstrated in isotopically purified samples depleted of nuclear spins, but in those cases the individual control of multiple nuclear-spin qubits is forgone.	
       
Here we realize a coherence time exceeding one second for a single electron spin in diamond that is coupled to a complex environment of multiple nuclear-spin qubits. We first use the electron spin as a quantum sensor to probe the microscopic structure of the surrounding nuclear-spin environment, including interactions between the nuclear spins. We find that the spin environment is accurately described by seven isolated single $^{13}C$ spins and six pairs of coupled $^{13}C$ spins (Fig. \ref{Figure1}a). We then develop pulse sequences to initialize, control and readout the state of the $^{13}C$-$^{13}C$ pairs. We use this control to directly characterize the coupling strength between the $^{13}C$ spins, thus revealing their atomic structure given by the distance between the two $^{13}C$ atoms and the angle they make with the magnetic field. Finally, we exploit this extensive knowledge of the microscopic environment to realize tailored decoupling sequences that effectively protect arbitrary quantum states stored in the electron spin for well over a second. This combination of a long electron spin coherence time and selective couplings to a system of up to 19 nuclear spins provides a promising path to multi-qubit registers for quantum sensing and quantum networks.\\


\noindent\textbf{RESULTS}

{\bf{System.}} We use a single nitrogen vacancy (NV) center (Fig. 1a) in a CVD-grown diamond at a temperature of 3.7 K with a natural $1.1 \%$ abundance of $^{13}$C and a negligible nitrogen concentration ($< 5$ parts per billion). A static magnetic field of $B_z \approx 403$ G is applied along the NV-axis with a permanent magnet (Methods). The NV electron spin is read out in a single shot with an average fidelity of $95 \% $ through spin-selective resonant excitation. The electron spin is controlled using microwave pulses through an on-chip stripline (Methods). \\ 

{\bf{Longitudinal relaxation.}} We first address the longitudinal relaxation ($T_1$) of the NV electron spin, which sets a limit on the maximum coherence time. At 3.7 Kelvin, spin-lattice relaxation due to two-phonon Raman and Orbach-type processes are negligible \cite{Takashi_PRL2008, Jarmola_PRL_2012}. No cross relaxation to P1 or other NV centers is expected due to the low nitrogen concentration. The electron spin can, however, relax due to microwave noise and laser background introduced by the experimental controls (Fig. 1). We ensure a high on/off ratio of the lasers ($> 100$ dB) and use switches to suppress microwave amplifier noise (see Methods). Figure \ref{Figure1}b shows the measured electron spin relaxation for all three initial states. We fit the average fidelity $F$ to 
\begin{equation}\label{T1}
F = 2/3e^{-t/T_1}+1/3.
\end{equation}
The obtained decay time $T_1$ is $(3.6 \pm 0.3)\cdot 10^3$ s. This value sets a lower limit for the spin-relaxation time, and is the longest reported for a single electron spin qubit.  Remarkably, the observed $T_1$ exceeds recent theoretical predictions based on single-phonon processes by more than an order of magnitude \cite{Astner_arxiv2017, Norambuena_arxiv2017}. To further investigate the origin of the decay, we prepare $m_s=0$ and measure the total spin population summed over all three states. The total population decays on a similar timescale ($\sim 3.6 \cdot 10^3$ s), indicating that the decay is caused by a reduction of the measurement contrast due to drifts in the optical setup, rather than by spin relaxation. This suggests that the spin-relaxation time significantly exceeds the measured $T_1$ value. Nevertheless, the long $T_1$ observed here already indicates that longitudinal relaxation is no longer a limiting factor for NV center coherence.\\


\begin{figure*}[p!]
\includegraphics[scale=0.8]{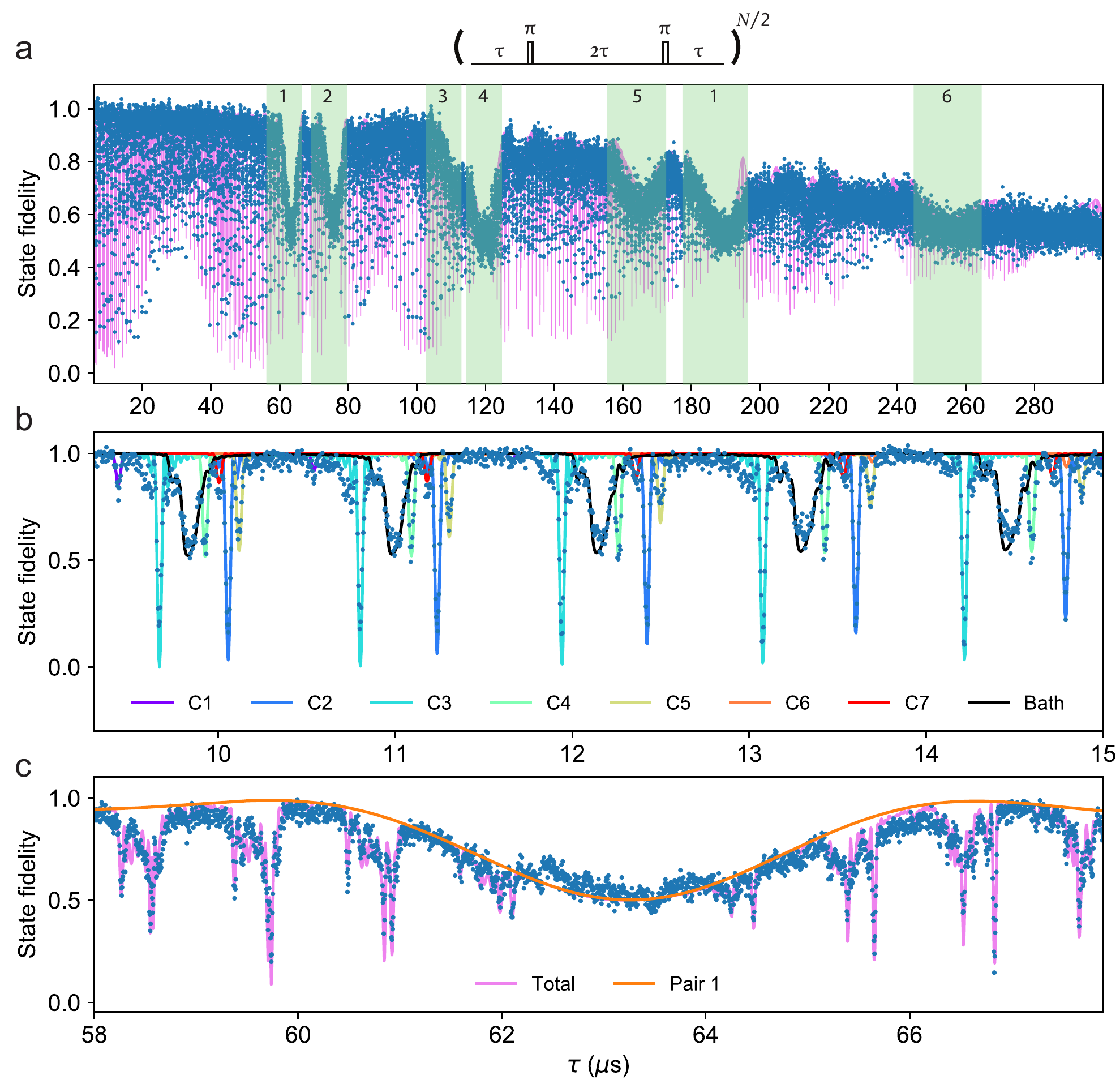}
\caption{\label{Figure2}
\textbf{Quantum sensing of the microscopic spin environment.} (a) Dynamical decoupling spectroscopy \cite{Taminiau_PRL2012} revealing a rich nuclear-spin environment consisting of individual $^{13}C$ spins, as well as pairs of coupled $^{13}C$ spins. The electron spin is prepared in a superposition, $|x\rangle = (|m_s = 0\rangle \pm|m_s = -1\rangle)/\sqrt{2}$ and a decoupling sequence of $N=32$ $\pi$-pulses separated by $2\tau$ is applied. Loss of coherence indicates the interaction of the electron spin with nuclear spins in the environment. Blue: data. Purple line: theory (see Methods). The shaded areas mark the signals due to six $^{13}C$-$^{13}C$ pairs labeled 1 to 6. (b) Zoom-in showing sharp signals due to coupling to isolated individual $^{13}C$ spins \cite{Taminiau_PRL2012, Kolkowitz_PRL2012, Zhao_NatureNano2012}. The total signal is well described by seven $^{13}C$ spins (see Supplementary Table 2 for hyperfine parameters) and a bath of 200 randomly generated spins with hyperfine couplings below 10 kHz. (c) Zoom-in showing a broad signal due to $^{13}C$-$^{13}C$ pair 1 \cite{Zhao_NatureNat2011,Shi_NatPhys2014}. Blue: data. The solid orange line is the theoretical signal just due to pair 1, while the purple line includes the seven individual $^{13}C$ spins and the $^{13}C$ spin bath as well.}
    \end{figure*}

{\bf{Quantum sensing of the microscopic spin environment.}} To study the electron spin coherence, we first use the electron spin as a quantum sensor to probe its nuclear-spin environment through dynamical decoupling spectroscopy \cite{Taminiau_PRL2012,Kolkowitz_PRL2012,Zhao_NatureNano2012}. The electron spin is prepared in a superposition $|x\rangle = (|m_s = 0\rangle + |m_s = -1\rangle)/\sqrt{2}$ and a dynamical decoupling sequence of $N$ $\pi$-pulses of the form $(\tau - \pi - \tau)^N$ is applied. The remaining electron coherence is then measured as a function of the time between the pulses $2\tau$. Loss of electron coherence indicates an interaction with the nuclear-spin environment.

The results in Fig. 2a for $N=32$ pulses reveal a rich structure consisting of both sharp and broader dips in the electron coherence. The sharp dips (Fig. 2b) have been identified previously as resonances due to the electron spin undergoing an entangling operation with individual isolated $^{13}C$ spins in the environment \cite{Kolkowitz_PRL2012,Taminiau_PRL2012, Zhao_NatureNano2012}. For this NV center, the observed signal is well explained by seven individual $^{13}C$ spins and a background bath of randomly generated $^{13}C$ spins (Fig. 2b). To verify this explanation we perform direct Ramsey spectroscopy on all seven spins (Supplementary Fig. 1) \cite{Cramer_NatureComm2016}. For the electron spin in $m_s = \pm 1$, each spin yields a single unique precession frequency due to the hyperfine coupling, indicating that all seven spins are distinct and do not couple strongly to other $^{13}C$ spins in the vicinity (See Supplementary Fig. 1).

The electron can be efficiently decoupled from the interactions with such isolated $^{13}C$ spins by setting $\tau = m \cdot \frac{2\pi}{\omega_L}$, with $m$ a positive integer and $\omega_L$ the $^{13}C$ Larmor frequency for $m_s=0$ \cite{Childress_Science2006}. In practice, however, this condition might not be exactly and simultaneously met for all spins due to: the limited timing resolution of $\tau$ (here 1 ns), measurement uncertainty in the value $\omega_L$, and differences between the $m_s=0$ frequencies for different $^{13}C$ spins, for example caused by different effective g-tensors under a slightly misaligned magnetic field (here $< 0.35^{\circ}$, Supplementary Note 3) \cite{Cramer_NatureComm2016,Childress_Science2006}. We numerically simulate these deviations from the ideal condition and find that, for our range of parameters, the effect on the electron coherence is small and can be neglected (Supplementary Fig. 2).     


\begin{figure*}[tb]
\includegraphics[scale=0.68]{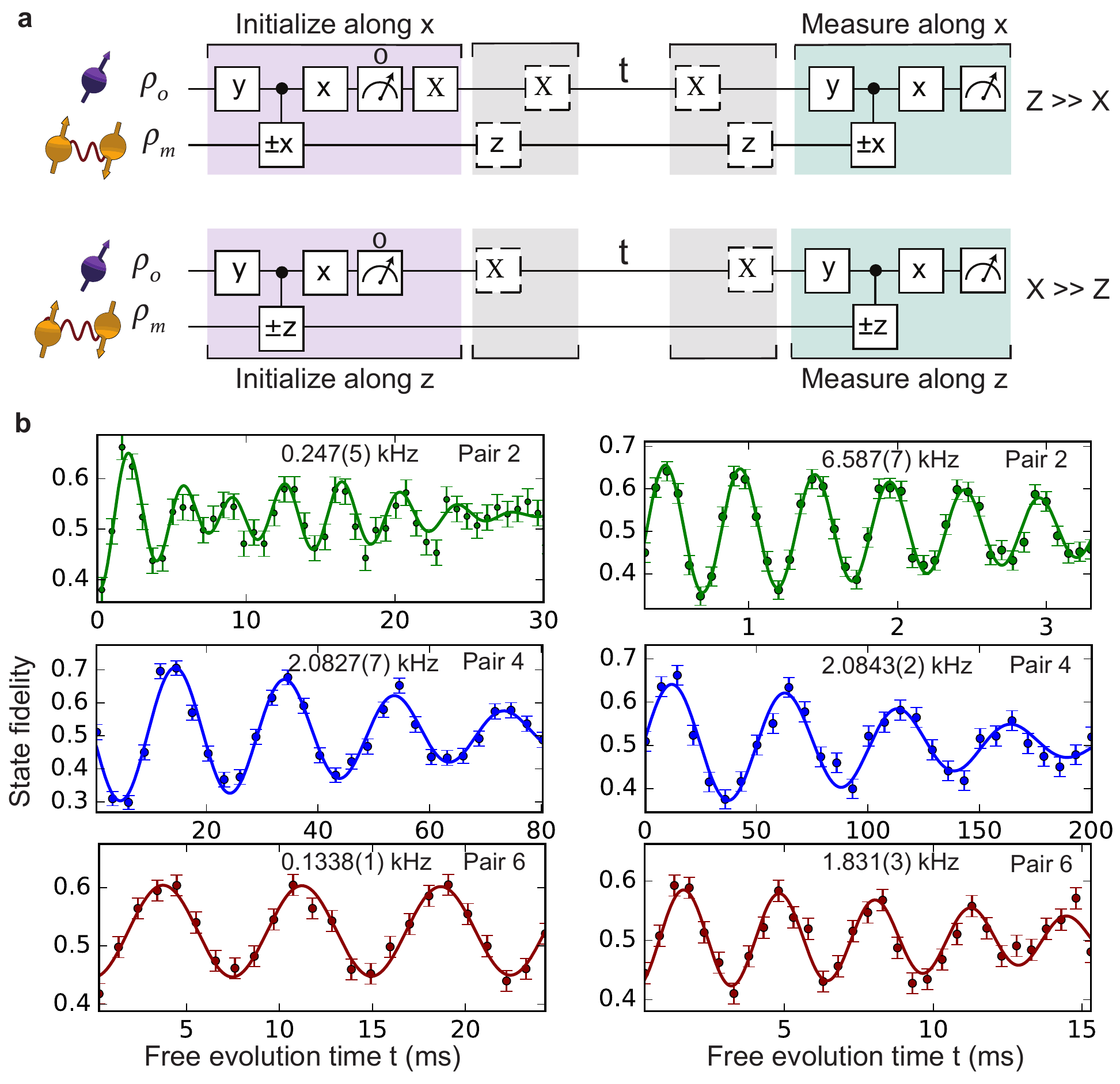}
\caption{\label{Figure3}
\textbf{Direct spectroscopy of nuclear-spin pairs.} (a) Measurement sequences for Ramsey spectroscopy of $^{13}C$ - $^{13}C$ pairs, for $Z >> X$ (top) and for $X >> Z$ (bottom). The controlled $\pm x$ ($ \pm z$) gates are controlled $\pm\pi/2$ rotations around $x$ ($z$) with the sign controlled by the electron state. (b) Nuclear spin Ramsey measurements for pairs 2, 4 and 6. The electron spin state during the free evolution time $t$ is set to $m_s=0$ (left) or $m_s=-1$ (right), and an artificial detuning is applied. Each pair yields a unique set of frequencies, confirming that the pairs are distinct. For pair 2 an additional beating is observed (frequency of $23(3)$ Hz), indicating a small coupling to one (or more) additional spins. See Supplementary Fig. 3 for the other three pairs and Supplementary Table 4 for fit results.}
\end{figure*}

\begin{figure*}[tb]
\includegraphics[scale=0.8]{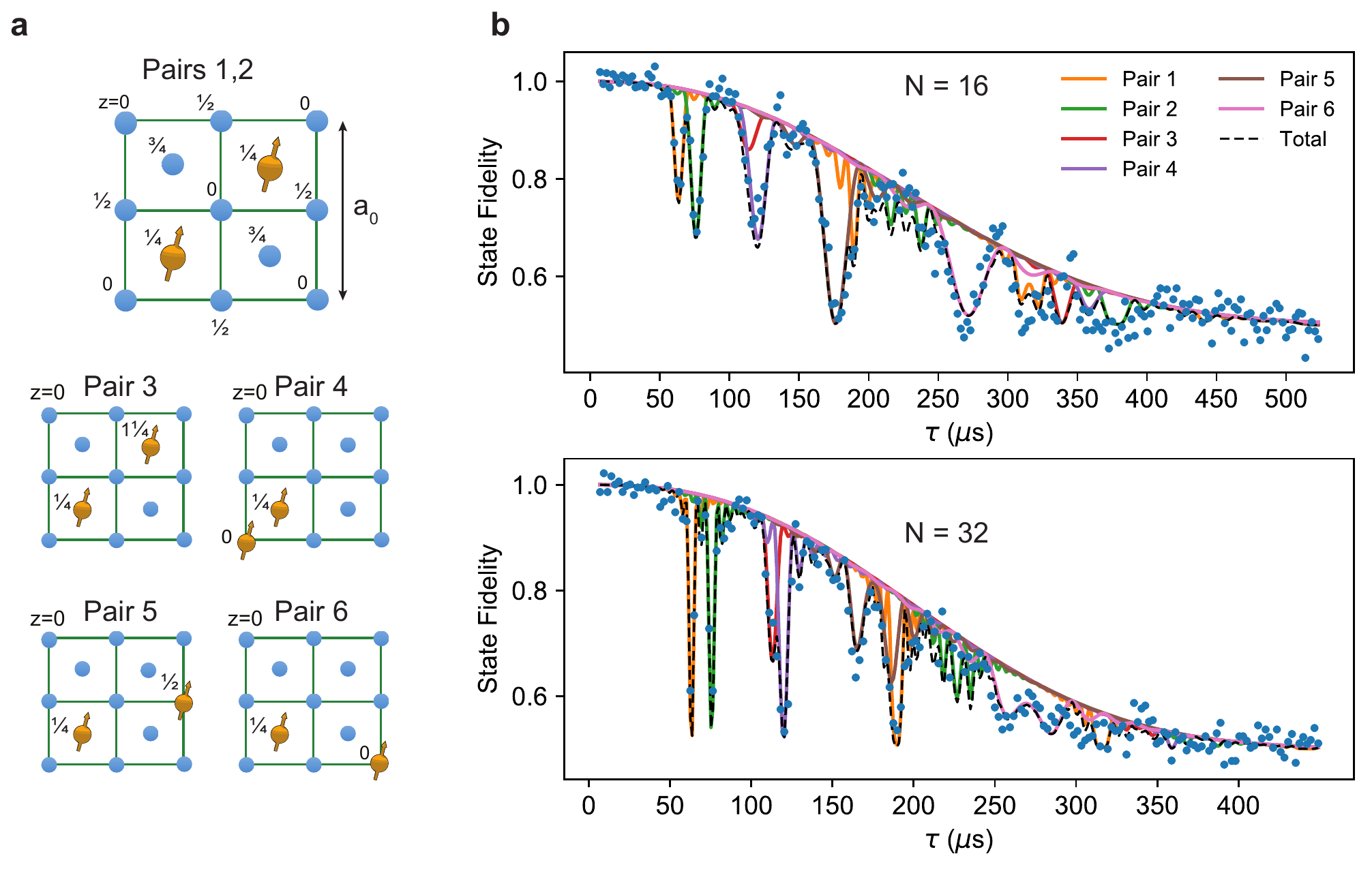}
\caption{\label{Figure4}
\textbf{Atomic structure and decoupling signal for the six nuclear spin pairs.} (a) Structure of the six $^{13}C$ - $^{13}C$ pairs within the diamond unit cell (up to rotational symmetries). The $z$ values give the height in fractions of the diamond lattice constant $a_0$. The magnetic field is oriented along the $<111>$ direction, i.e. along the axis of pair 4. (b) The calculated signal for the six individual $^{13}C$ - $^{13}C$ pairs accurately describes the measured decoupling signal for different number of pulses $N$. Data is taken for $\tau = m\cdot \frac{2\pi}{\omega_L}$ to avoid coupling to single $^{13}C$ spins. See Supplementary Fig. 4 for other values of $N$.}
\end{figure*}

\begin{figure*}[tb]
\includegraphics[scale=0.8]{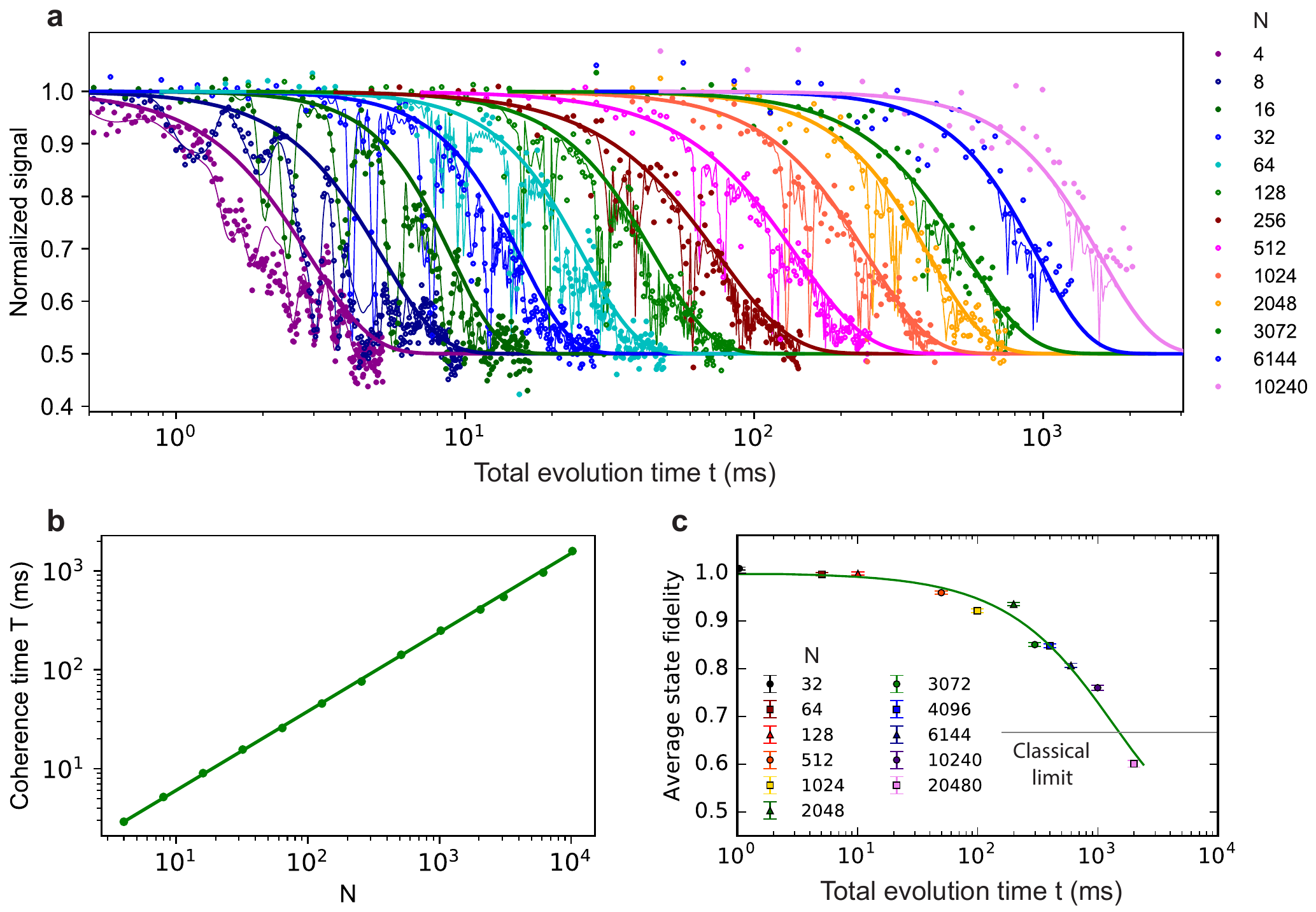}
\caption{\label{Figure5}
\textbf{Protecting quantum states with tailored decoupling sequences.} (a) Normalized signal under dynamical decoupling with the number of pulses varying from $N=4$ to $N=10240$. The electron is initialized and readout along $x$. The thin lines are fits to equation (\ref{eqn:fit_pairs}), which takes into account the six identified $^{13}C-^{13}C$ pairs. We use the extracted amplitudes $A$ to re-normalize the signal. Thick lines are the extracted envelops ($0.5+ 0.5 \cdot e^{-(t/T)^{n}} $) with $T$ and $n$ obtained from the fits. See Supplementary Fig. 5 for the obtained values $n$. (b) Scaling of the obtained coherence time $T$ as function of the number of pulses (error bars are $ < 5 \%$). The solid line is a fit to the power function $T_{N=4}\cdot (N/4)^\eta$, where $T_{N=4}$ is the coherence time for $N=4$. We find $\eta=0.799 (2)$. (c) The average state fidelity obtained for the six cardinal states (Supplementary Fig. 6). Unlike in (a), the signal is shown without any renormalization. The number of pulses $N$ is chosen to maximize the obtained signal at the given total evolution time while avoiding interactions with the $^{13}C$ environment. The solid green line is a fit to an exponential decay. The horizontal line at $\frac{2}{3}$ fidelity marks the classical limit for storing quantum states. The two curves cross at $t=1.46$ s demonstrating the protection of arbitrary quantum states well beyond a second.}
\end{figure*}

We associate the broader dips in Fig. 2a and Fig. 2c to 
pairs of strongly coupled $^{13}C$ spins. Such $^{13}C$ - $^{13}C$ pairs were treated theoretically \cite{Zhao_NatureNat2011,Wang_NatComm2017} and the signal due to a single pair of nearest-neighbor $^{13}C$ spins with particularly strong couplings to a NV center has been detected \cite{Shi_NatPhys2014}. In this work, we exploit improved coherence times to detect up to six pairs, including previously undetected non-nearest-neighbor pairs. We then develop pulse sequences to polarize and coherently control these pairs to be able to directly reveal their atomic structure through spectroscopy.\\

{\bf{Direct spectroscopy of nuclear-spin pairs.}} The evolution of $^{13}C$ - $^{13}C$ pairs can be understood from an approximate pseudo-spin model in the subspace spanned by $\ket{\uparrow\downarrow} = \ket{\Uparrow }$ and $\ket{\downarrow\uparrow} = \ket{\Downarrow }$, following Zhao et al. \cite{Zhao_NatureNat2011} (Supplementary Notes 1 and 2). The pseudo-spin Hamiltonian depends on the electron spin state. For $m_s=0$ we have: 
\begin{equation}\label{PseudoSpinH0}
\hat{H}_0 = X\hat{S_x}
\end{equation}
and for $m_s = -1$: 
\begin{equation}\label{PseudoSpinH1}
\hat{H}_1 = X\hat{S_x} + Z\hat{S_z}
\end{equation}
with $\hat{S_x}$ and $\hat{S_z}$ the Pauli spin operators. $X$ is the dipolar coupling between the $^{13}C$ spins and $Z$ is due to the hyperfine field gradient (see Methods) \cite{Zhao_NatureNat2011}. The evolution of the $^{13}C$ - $^{13}C$ pair during a decoupling sequence will thus in general depend on the initial electron spin state, causing a loss of electron coherence. 

We now show that this conditional evolution enables direct spectroscopy of the $^{13}C$ - $^{13}C$ interaction $X$. Consider two limiting cases: $X >> Z$ and $Z >> X$, which cover the pairs observed in this work. In both cases, loss of the electron coherence is expected for the resonance condition $\tau = \tau_k = (2k-1)\frac{\pi}{2\omega}$, with $k$ a positive integer and $\omega = \sqrt{X^2+(Z/2)^2}$ \cite{Zhao_NatureNat2011, Taminiau_PRL2012, Sar_Nature2012}. For $X >> Z$ the net evolution at resonance is a rotation around the $z$-axis with the rotation direction conditional on the initial electron state (mathematically analogous to the case of a single $^{13}C$ spin in a strong magnetic field \cite{Taminiau_PRL2012, Taminiau_NatNano2014}). For $Z >> X$ the net evolution is a conditional rotation around the $x$-axis (analogous to the Nitrogen nuclear spin subjected to a driving field \cite{Sar_Nature2012}). These conditional rotations provide the controlled gate operations required to initialize, coherently control and directly probe the pseudo-spin states.

The measurement sequences for the two cases are shown in Fig. 3a. First, a dynamical decoupling sequence is performed that correlates the electron state with the pseudo-spin state. Reading out the electron spin in a single shot then performs a projective measurement that prepares the pseudo-spin into a polarized state. For $X >> Z$ the pseudo-spin is measured along its $z$-axis and thus prepared in $\ket{\Uparrow}$. For $Z >> X$ the measurement is along the $x$-axis and the spin is prepared in $(\ket{\Uparrow} + \ket{\Downarrow})/\sqrt{2}$. Second, we let the pseudo-spin evolve freely with the electron spin in one of its eigenstates ($m_s=0$ or $m_s=-1$) so that we directly probe the precession frequencies $\omega_0 = X$ or $\omega_1 = \sqrt{X^2+Z^2}$. For $Z >> X$, an extra complication is that the initial state $(\ket{\Uparrow} + \ket{\Downarrow})/\sqrt{2}$ is an eigenstate of $\hat{H}_0$. To access $\omega_0 = X$, we prepare $(\ket{\Uparrow}+ i\ket{\Downarrow})/\sqrt{2}$ - a superposition of $\hat{H}_0$ eigenstates - by first letting the system evolve under $\hat{H}_{1}$ for a time $\pi/(2\omega_1)$. Finally the state of the pseudo-spin is read out through a second measurement sequence.

We find six distinct sets of frequencies (Fig. 3b), indicating that six different $^{13}C$ - $^{13}C$ pairs are detected. The measurements for $m_s=0$ directly yield the coupling strengths $X$ and therefore the atomic structure of the pairs (Fig. 4a). We observe a variety of coupling strengths corresponding to nearest-neighbor pairs ($X/2\pi = 2082.7(7)$  Hz, theoretical value $2061$ Hz) as well as pairs separated by several bond lengths (e.g. $X/2\pi = 133.8(1)$ Hz, theoretical value $133.4$ Hz). Note that for pair 4 we have $X >> Z$, so the resonance condition is mainly governed by the coupling strength $X$. This makes it likely that additional pairs with the same $X$ \textemdash\ but smaller $Z$ values \textemdash\ contribute to the observed signal at $\tau = 120\ \mu$s. Nevertheless, the environment can be described accurately by the six identified pairs, which we verify by comparing the measured dynamical decoupling curves for different values of $N$ to the calculated signal based on the extracted couplings (Fig. 4b).\\ 

{\bf{Electron spin coherence time.}} Next, we exploit the obtained microscopic picture of the nuclear spin environment to investigate the electron spin coherence under dynamical decoupling. To extract the loss of coherence due to the remainder of the dynamics of the environment, i.e. excluding the identified signals from the $^{13}C$ spins and pairs, we fit the results to:
\begin{equation}
F = \frac{1}{2} + A \cdot M(t)\cdot e^{-(t/T)^{n}},
\label{eqn:fit_pairs}
\end{equation}
in which $M(t)$ accounts for the signal due to the coupling to the $^{13}C$-$^{13}C$ pairs (Fig. 4b, see Methods). $A$, $T$ and $n$ are fit parameters that account for the decay of the envelope due to the rest of the dynamics of the environment and pulse errors. As before, interactions with individual $^{13}C$ spins are avoided by setting $\tau = m\cdot\frac{2\pi}{\omega_L}$. An additional challenge is that at high numbers of pulses the electron spin becomes sensitive even to small effects, such as spurious harmonics due to finite MW pulse durations \cite{PhysRevX.5.021009,PhysRevApplied.7.054009} and non-secular Hamiltonian terms \cite{ajoy2016dc}, which cause loss of coherence over narrow ranges of $\tau$ ($ < 10$ ns). Here, we avoid such effects by scanning a range of $\sim20$ ns around the target value to determine the optimum value of $\tau$.

Figure 5a shows the electron coherence for sequences from $N=4$ to $10240$ pulses. The coherence times $T$, extracted from the envelopes, reveal that the electron coherence can be greatly extended by increasing the number of pulses $N$. The maximum coherence time is $T= 1.58(7)$ seconds for $N=10240$ (Fig. 5b). We determine the scaling of $T$ with $N$ by fitting to $T_{N=4} \cdot (N/4)^\eta$, with $T_{N=4} $ the coherence time for $N=4$ \cite{Lange_science2010,Ryan_PRL2010,naydenov2011dynamical,Medford_PRL2012, Bar-Gill_NatComm2012} which gives $\eta=0.799(2)$. No saturation of the coherence time $T$ is observed yet, so that longer coherence times are expected to be possible. In our experiments, however, pulse errors become the limiting factor at larger $N$, causing a decrease in the amplitude $A$. 


{\bf{Protecting arbitrary quantum states.}} Finally, we demonstrate that arbitrary quantum states can be stored in the electron spin for well over a second by using decoupling sequences that are tailored to the specific microscopic spin environment (Fig. 5c). For a given storage time, we select $\tau$ and $N$ to maximize the obtained fidelity by avoiding interactions with the characterized $^{13}C$ spins and $^{13}C$-$^{13}C$ pairs. To asses the ability to protect arbitrary quantum states, we average the storage fidelity over the six cardinal states and do not re-normalize the results. The results show that quantum states are protected with a fidelity above the $2/3$ limit of a classical memory for at least 0.995 seconds (using $N=10240$ pulses) and up to $1.46$ seconds from interpolation of the results. These are the longest coherence times reported for single solid-state electron spin qubits \cite{Muhonen_NatNano2014}, despite the presence of a dense nuclear spin environment that provides multiple qubits. \\ 


\noindent\textbf{DISCUSSION}

These results provide new opportunities for quantum sensing and quantum information processing, and  are applicable to a wide variety of solid-state spin systems \cite{Seo_NatComm2016, Widmann_NatMat2014, Falk_PRL2015, Yang_PRB2014, Rogers_PRL2014, Sukachev_Arxiv2017, Becker_Arxiv2017, Rose_Arxiv2017, Siyushev_PRB2017, Pla_PRL2014, Muhonen_NatNano2014, Wolfowicz_NJP2016, Lee_NatNano2013, Iwasaki_arXiv2017}. First, these experiments are a proof-of-principle for resolving the microscopic structure of multi-spin systems, including the interactions between spins \cite{Zhao_NatureNat2011}. The developed methods might be applied to detect and control spin interactions in samples external to the host material \cite{Lovchinsky_Science2016,kucsko2013nanometer,shi2015single,tetienne2014nanoscale}. Second, the combination of long coherence times and selective control in an electron-nuclear system containing up to twenty spins enables improved multi-qubit quantum registers for quantum networks. The electron spin coherence now exceeds the time needed to entangle remote NV centers through a photonic link, making deterministic entanglement delivery possible \cite{Humphreys2017}. Moreover, the realized control over multiple $^{13}C$-$^{13}C$ pairs provides promising new multi-qubit quantum memories with long coherence times, as the pseudo-spin naturally forms a decoherence-protected subspace \cite{Lidar_PRL1998}. \\    


\noindent\textbf{METHODS}

\textbf{Setup.} The experiments are performed at $3.7$ Kelvin (Montana Cryostation) with a magnetic field of $\sim$403 G applied along the NV axis by a permanent magnet. We realize long relaxation ($T_1 > 1$ hour) and coherence times ($ > 1$ second) in combination with fast spin operations (Rabi frequency of 14 MHz) and readout/initialization ($\sim 10\ \mu$s), by minimizing noise and background from the microwave (MW) and optical controls. Amplifier (AR 25S1G6) noise is suppressed by a fast microwave switch (TriQuint TGS2355-SM) with a suppression ratio of ~40 dB. Video leakage noise generated by the switch is filtered with a high pass filter. We use Hermite pulse envelopes \cite{Lieven,warren1984effects} to obtain effective MW pulses without initialization of the intrinsic $^{14}N$ nuclear spin. To mitigate pulse errors we alternate the phases of the pulses following the $XY8$ scheme \cite{XY8}. Laser pulses are generated by direct current modulation (515 nm laser, Cobolt MLD - for charge state control) or by acoustic optical modulators (637 nm Toptica DL Pro and New Focus TLB-6704-P â for spin pumping and single-shot readout \cite{Robledo_Nature2011}). The direct current modulation yields an on/off ratio of $>135$ dB. By placing two modulators in series (Gooch and Housego Fibre Q) an on/off ratio of $> 100$ dB is obtained for the 637 nm lasers.    

\textbf{Sample.} We use a naturally occurring Nitrogen-Vacancy (NV) center in high-purity type IIa homoepitaxially chemical-vapor-deposition (CVD) grown diamond with a $1.1\%$ natural abundance of $^{13}$C and a $\langle 111 \rangle$ crystal orientation (Element
Six). To enhance the collection efficiency a solid-immersion lens was fabricated on top of the NV center \cite{Robledo_Nature2011,Hadden_APL2010} and a single-layer aluminum-oxide anti-reflection coating was deposited \cite{Pfaff_Science2014,Yeung_APL2012}.  

\textbf{Data analysis.} We describe the total signal for the NV electron spin after a decoupling sequence in Fig. 2 as:
\begin{equation}
F = \frac{1}{2} + A \cdot M_{bath}(t)\cdot \prod_{i=1}^{7} M_C^{i}(t) \cdot \prod_{j=1}^{6} M_{pair}^{j}(t) \cdot e^{-(t/T)^{n}},
\label{eqn:signal_total}
\end{equation}
where $t$ is the total time. $M_{bath}$ is the signal due to a randomly generated background bath of non-interacting spins with hyperfine couplings below 10 kHz. $M_{C}^{i}$ are the signals due to the seven individual isolated $^{13}C$ spins \cite{Taminiau_PRL2012}. $M_{pair}^{j}$ are the signals due to the six $^{13}C-^{13}C$ pairs and are given by $1/2+Re (Tr(U_0U_1^{\dag}))/4$, $U_0$ and $U_1$ the evolution operators of the pseudo-spin pair for the decoupling sequence conditional on the initial electron state ($m_s = 0$ or $m_s = -1$) \cite{Zhao_NatureNat2011}. The coherence time $T$ and exponent $n$ describe the decoherence due to remainder of the dynamics of the spin environment. 

Setting $\tau = m \cdot 2\pi / \omega_L$ avoids the resonances due to individual $^{13}C$ spins, so that equation (\ref{eqn:signal_total}) reduces to:
\begin{equation}
F = \frac{1}{2} + A \cdot \prod_{j=1}^{6} M_{pair}^{j}(t) \cdot e^{-(t/T)^{n}}.
\label{eqn:signal_pairs}
\end{equation}
The data in Fig. \ref{Figure4} and \ref{Figure5} are fitted to equation (\ref{eqn:signal_pairs}) and $A$, $T$ and $n$ are extracted from these fits. \\

\noindent\textbf{Acknowledgements}\\
We thank V. V. Dobrovitski, J. E. Lang, T. S. Monteiro, C. L. Degen, and R. Hanson for valuable discussions, P. Vinke, R. Vermeulen, R. Schouten and M. Eschen for help with the experimental apparatus, and A. J. Stolk for characterization measurements. We acknowledge support from the Netherlands Organization for Scientific Research (NWO) through a Vidi grant.\\
\noindent\textbf{Author contributions}\\
MHA and THT devised the experiments. MHA, JC and THT constructed the experimental apparatus. MM and DJT grew the diamond. MHA performed the experiments with support from MAB and NK. MHA and THT analyzed the data with help of all authors. THT supervised the project.



\clearpage


\widetext
\begin{center}
\textbf{\large Supplementary Information for ``One-second coherence for a single electron spin coupled to a multi-qubit nuclear-spin environment''}
\end{center}

\newcommand{\angstrom}{\textup{\AA}}
\renewcommand{\figurename}{\textbf{Supplementary Figure}}
\renewcommand{\tablename}{\textbf{Supplementary Table }}

\setcounter{equation}{0}
\setcounter{figure}{0}
\setcounter{table}{0}
\setcounter{page}{1}
\makeatletter
\renewcommand{\theequation}{S\arabic{equation}}

\section{SUPPLEMENTARY NOTE 1: System Hamiltonian.}
 The Hamiltonian describing a system composed of an NV center and a $^{13}C$ nuclear spin environment, in a suitable rotating frame and under the secular approximation can be described by:
\begin{equation}
\hat{H} =  \sum_{i=1}^{n} (\omega_0\hat{I}_{z}^{i} + A_{\parallel}^{i} \hat{S}_{z}\hat{I}_{z}^{i} + A_{\perp}^{i} \hat{S}_{z} \hat{I}_{x}^{i} ) + \hat{H}_{n-n},
\label{eq:system_Hamiltonian}
\end{equation}
where $\omega_o$ ($= 2\pi \cdot \gamma_c B_z$) is the bare Larmor frequency, $A_{\parallel}(A_{\perp})$ is the parallel (perpendicular) hyperfine coupling between the electron and $^{13}C$ nuclear spin with respect to the applied static magnetic field. The dipolar interaction between $^{13}C$ nuclear spins in the environment $H_{n-n}$ is given by:

\begin{equation}
\hat{H}_{n-n} = \sum_{i>j} \frac{\mu_o}{4\pi} \frac{\gamma_c^{i}\gamma_c^{j}}{r_{ij}^{3}} [\mathbf{I}^i \cdot \mathbf{I}^{j}  - 3(\mathbf{I}^i \cdot \hat{r}_{ij})(\mathbf{I}^j\cdot \hat{r}_{ij})],
\end{equation}
where $\gamma_c$ is the gyromagnetic ratio of the nuclear spin, $\hat{r}_{ij}$ is the unit vector connecting the two nuclear spins and $ \mathbf{I}^{i}$ is the Pauli spin operator. Now we can rewrite the Hamiltonian as follows:

\begin{align}
	\hat{H} = &\ket{0}\bra{0}\hat{H}_{0} + \ket{1}\bra{1}\hat{H}_{1},\\
    \hat{H_{0}} = & \hat{H}_{n-n} + \sum_{i=1}^{n} \omega_0 \hat{I^{i}_{z}},\\
	\hat{H_{1}} = & \hat{H}_{n-n} + \sum_{i=1}^{n} (\omega_0 - A_{\parallel}^{i}) \hat{I}^{i}_{z} + A_{\perp}^{i} \hat{I}^{i}_{x},
\end{align}
where $H_{0}$ ($H_{1}$) is the Hamiltonian describing the rest of the system if the electron is in the state $m_s = 0$ ($m_s = -1$).

\section{supplementary note 2: Pseudo-spin model}
Under high magnetic field, the dynamics of a $^{13}C-^{13}C$ pair can be approximated by a pseudo-spin-$\frac{1}{2}$ model \cite{Zhao_NatureNano2012_S,Shi_NatPhys2014_S}, where the two anti-parallel spin states of the pair ($\ket{\uparrow\downarrow}$ and $\ket{\downarrow\uparrow}$) are mapped into spin-up ($\ket{\Uparrow }$) and spin-down ($\ket{\Downarrow }$) states of the pseudo-spin. The polarized states ($\ket{\uparrow\uparrow}$ and $\ket{\downarrow\downarrow}$) have large energy separation (due to large Zeeman energy) with respect to other states and thus do not play a role in the dynamics. Under these assumptions, the dynamics of the pseudo-spin can be described conditional on the electron spin state by the Hamiltonian: 
\begin{equation}
\hat{H}_0 = X\hat{S_x}, \textnormal{ and    $\hat{H}_1 = X\hat{S_x} + Z\hat{S_z}$,}  
\end{equation}
where $\hat{H}_0 $ ($\hat{H}_1$) is the Hamiltonian if the electron is in $m_s = 0$ ($m_s = -1$), $X$ is the dipolar coupling strength between the two nuclear spins \cite{Zhao_NatureNat2011_S}:
\begin{equation}
X =  \dfrac{\mu_0 }{4\pi}  \dfrac{\gamma_c^2}{r^3} \dfrac{1}{2} (1-3 \cos^2 \theta ),
\label{eq:coupling_strength}
\end{equation}
where $\gamma_c$ is the gyromagnetic ratio of the $^{13}C$ nuclear spin, $r$ is the distance between the two nuclear spins forming the pair, $\theta$ is the angle between the pair axis $\vec{r}$ and the external magnetic field direction ([1,1,1] in our case). Z  is due to the hyperfine field gradient \cite{Shi_NatPhys2014_S},

\begin{equation}
Z = Z_{\parallel}+Z_{\perp} = (A_{\parallel}^{1}-A_{\parallel}^{2})+\frac{(A_{\perp}^{1})^{2}-(A_{\perp}^{2})^{2}}{2\pi \cdot \gamma_C B_0}. 
\end{equation}

 Supplementary Table 3 shows the calculated set of possible coupling strengths for pairs with different distances, $r$, and angles, $\theta$, starting from the most strongly coupled pair (2.061 kHz) down to a coupling strength of 76 Hz. This is the range of interest for the pairs that we detect in this work. The values of the coupling strength $X$ are significantly different for different possible pairs and thus enable us to determine the distance between the two nuclear spins forming the pair and their orientation with respect to the external field. Although this information is enough to describe the dynamics under dynamical decoupling and the electron spin coherence, the measured value of $Z$ for a single electron-spin state, i.e. $m_s = -1$ in our case, does not yet enable us to uniquely determine the relative position of the pair with respect to the NV. Measuring $Z$ for $m_s = +1$ as well enables obtaining the two quantities ($A_{\parallel}^{1}-A_{\parallel}^{2}$) and ($A_{\perp}^{1})^{2}-(A_{\perp}^{2})^{2}$, which further narrows down the possible pair positions \cite{Shi_NatPhys2014_S}. 

\section{SUPPLEMENTARY NOTE 3: Effect of magnetic field misalignment on $^{13}C$ precession frequencies}

A misaligned field from the NV-axis would give rise to non-secular terms in the Hamiltonian leading to an effective g-tensor for $^{13}C$ nuclear spins that depends on the hyperfine coupling strength between the electron and the $^{13}C$ nuclear spin. For the electron in $m_s = 0$, this effective g-tensor can be calculated as follows \citep{Childress_Science2006_S}:

\begin{gather}
\hat{g}(m_s = 0) 
= 
 \begin{bmatrix} 1+\eta A_{xx} & \eta A_{xy} & \eta A_{xz} \\ \eta A_{xy} & 1+ \eta A_{yy} & \eta A_{yz} \\ 0 & 0 &1 
 \end{bmatrix} ,
\end{gather}
where $\eta = \frac{2 \gamma_e}{2\pi \cdot \gamma_c \Delta } = \frac{1}{2 \pi} \cdot 1.824 \cdot 10^{-3}$ kHz$^{-1}$, and $A_{mn}$ is the hyperfine tensor between the electron and $^{13}C$ nuclear spin. The bare $^{13}C$ precession frequency can now be calculated as $\omega_0 = \abs{2\pi \gamma_c \vec{B} \cdot \hat{g}(0)} $ \citep{Childress_Science2006_S}. We estimate our magnetic field alignment to be better than  $0.35$ degrees, which corresponds to a maximum perpendicular field component of 2.5 G ( see Supplementary Table \ref{Tab:ExpParams}). Now if we assume that our field lies in $xz$-plane, i.e. $\vec{B} = B_z \hat{e_z} + B_x \hat{e_x}$, then $\vec{B} \cdot \hat{g} = B_x (1+\eta A_{xx}) \hat{e_x} + \eta B_x A_{xy} \hat{e_y}+ (\eta B_x A_{xz} + B_{z}) \hat{e_z}$, which leads to :
\begin{equation}
\omega_0 = 2\pi \cdot \gamma_c \sqrt{[B_x(1+\eta A_{xx})]^2 + [\eta B_x A_{xy}]^2 +[B_z + \eta B_x A_{xz}]^2 }
\end{equation}

In our case $A_{xz}$ and $A_{xx}$ range from $-2\pi \cdot 50$ to $2\pi \cdot 50$ kHz (at maximum), and the maximum value of $B_{x}$ is 2.5 G. This means that different nuclear spins would have different bare Larmor precession frequencies, $\omega_0$, depending on their hyperfine coupling parameters. The dominant term of change in $\omega_0$ with the hyperfine coupling strengths, for our range of parameters, is $2\pi \cdot \gamma_c \eta B_x A_{xz}$, which would lead to a maximum difference in $\omega_0$ of $2\pi \cdot 500$ Hz between different nuclear spins. This is consistent with what we experimentally observe (see Supplementary Table \ref{Tab:Spins}). 
\clearpage

\begin{figure*}[tb]
\centering
\includegraphics[scale=0.8]{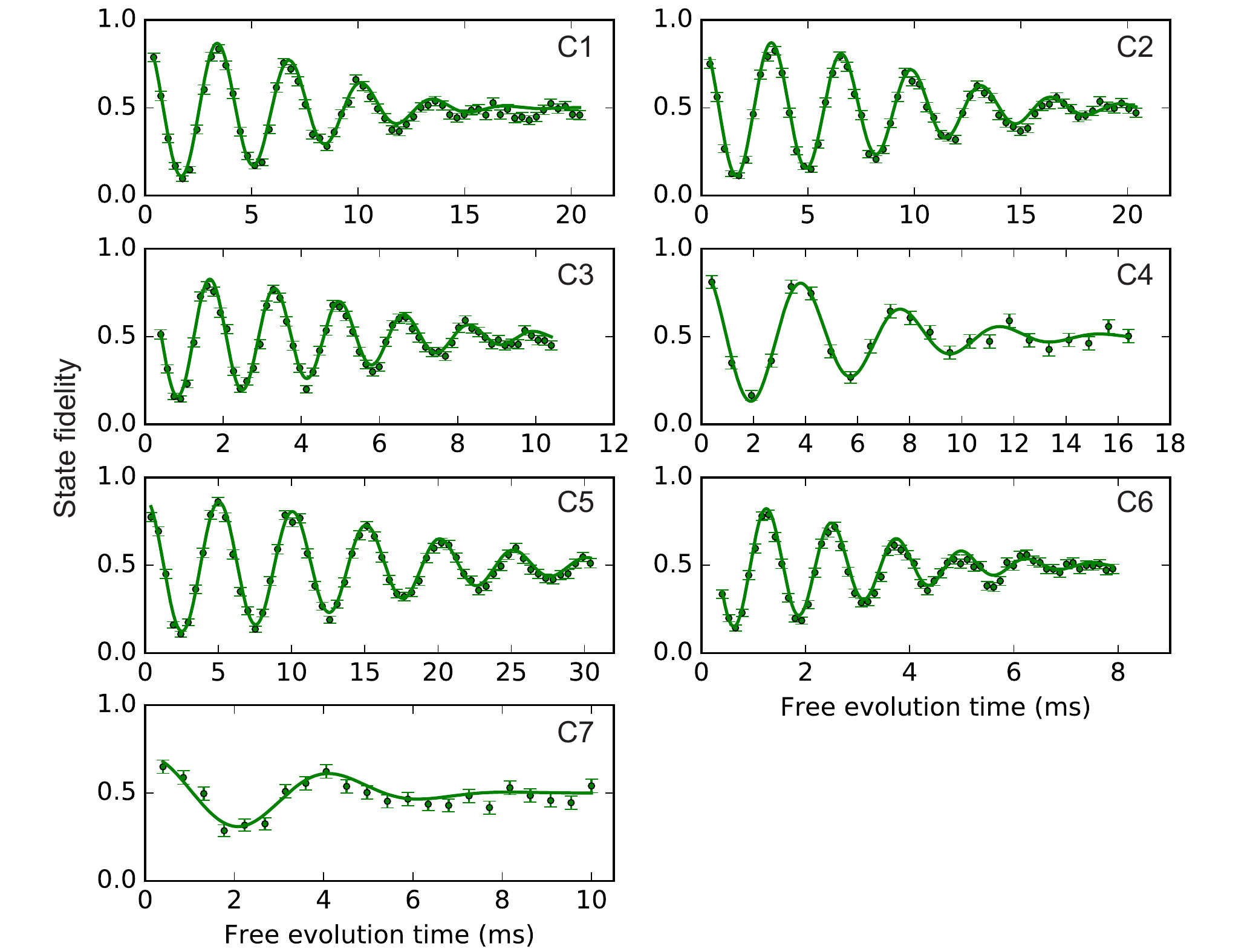}
\caption{\label{Figure 5}
\textbf{Ramsey experiments for the seven individual $^{13}C$ spins.} Ramsey interferometry \cite{Cramer_NatureComm2016_S} for the $^{13}C$ nuclear spins. The electron spin state during the free evolution time is $m_s = -1$ ($+1$ for C4 and C7). Lines are sinusoidal fits with a Gaussian decay: $F  = a + A\cdot e^{-(t/T_2^*)^2}\cos{(\delta t+ \phi)}$, with $t$ the free evolution time and $\delta$ a detuning. All seven spin signals are well described by a single, unique precession frequency $\omega_{\pm1} \approx \omega_0 \pm A_{\parallel}$ (see Supplementary Table \ref{Tab:Spins}) and a Gaussian decay, indicating that all seven spins are distinct and that none couple strongly to other $^{13}C$ spins in the environment. The minimum coupling strength for the observed $^{13}C$-$^{13}C$  pairs of $83$ Hz (Supplementary Table \ref{Tab:Clusters}), would already introduce a clear beating in $\sim 3$ ms, indicating that the seven identified single $^{13}C$ spins are not part of the $6$ detected $^{13}C$ - $^{13}C$ pairs.}
	\end{figure*}

\begin{figure*}[tb]
\centering
\includegraphics[scale=0.8]{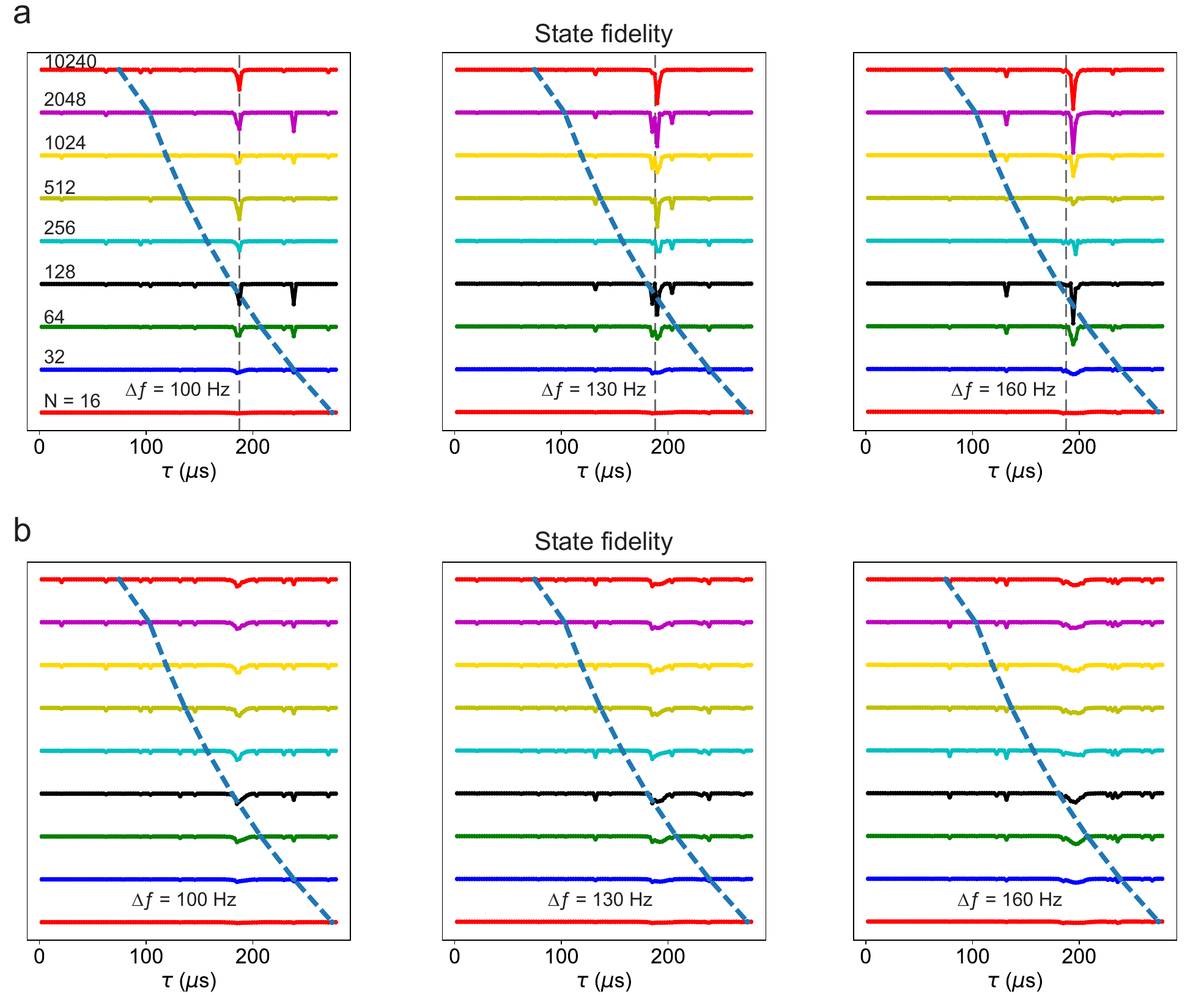}
\caption{\label{Figure_delta_f}
\textbf{Effect of deviations of $\tau$ from $2\pi/\omega_0$.} To avoid coupling to the single $^{13}$C spins, we aim to set the interpulse spacing to the revival condition $\tau = m\cdot2\pi/\omega_0$, with $m$ an integer and $\omega_0$ the $^{13}C$ frequency for $m_s=0$ \cite{Childress_Science2006_S}. However, this condition is not met exactly and simultaneously in the experiments for all $^{13}$C spins (Supplementary Note 3 and Supplementary Table \ref{Tab:Spins}). Here we explore the effect of small deviations from this condition. \textbf{a)} Simulated electron spin fidelity after a decoupling sequence with $\tau = m \cdot \frac{2\pi}{\omega_L}$, with $\omega_L$ the $^{13}C$ Larmor frequency estimated from ESR measurements (Supplementary Table \ref{Tab:ExpParams}). In these simulations we include the seven characterized $^{13}C$ spins and set all precession frequencies to $\omega_0 = \omega_L-\Delta \omega$.  The curves show results for $\frac{\Delta \omega}{2\pi} = 100, 130$ and $160$ Hz. The $y$-axis scale is such that the difference between horizontal lines at $\tau = 0$ is $1$. The dashed blue line marks the $1/e$ decay times for different values of N (from Fig. 5); the main region of interest lies to the left of this line. The vertical gray dashed line provides a visual aid to illustrate how the dip positions change with $\Delta \omega$. This shows that a change of 30 Hz in $\frac{\Delta \omega}{2\pi}$ leads to variations of the dip pattern. \textbf{b)} The obtained state fidelity averaged over 500 repetitions with $\omega_0$ for the seven spins drawn from a Gaussian distribution with a mean frequency of $\omega_L - \Delta \omega$ and standard deviation of $30$ Hz. These fluctuations match the typical observed values of ${T_{2}}^{*}$ for the nuclear spins. The result shows that differences in $m_s=0$ frequencies for different $^{13}C$ spins are smeared out by dephasing, so that their net effect on the decoupling curves is small. Additionally, the interpulse delay is set with a precision of $\delta_p = 1$ ns. The maximum relative error occurs at short $\tau$ ($\tau=\tau_L$) and is of order $\delta_p/2\tau$. This is equivalent to a $\Delta\omega \sim  2\pi \cdot 100\ $Hz, which the simulations show has a negligible effect at short $\tau$. At larger $\tau$ ($\tau > 10\tau_L$) the relative error in $\tau$ quickly becomes negligible.}
    \end{figure*}

\begin{figure*}[tb]
\centering
\includegraphics[scale=0.80]{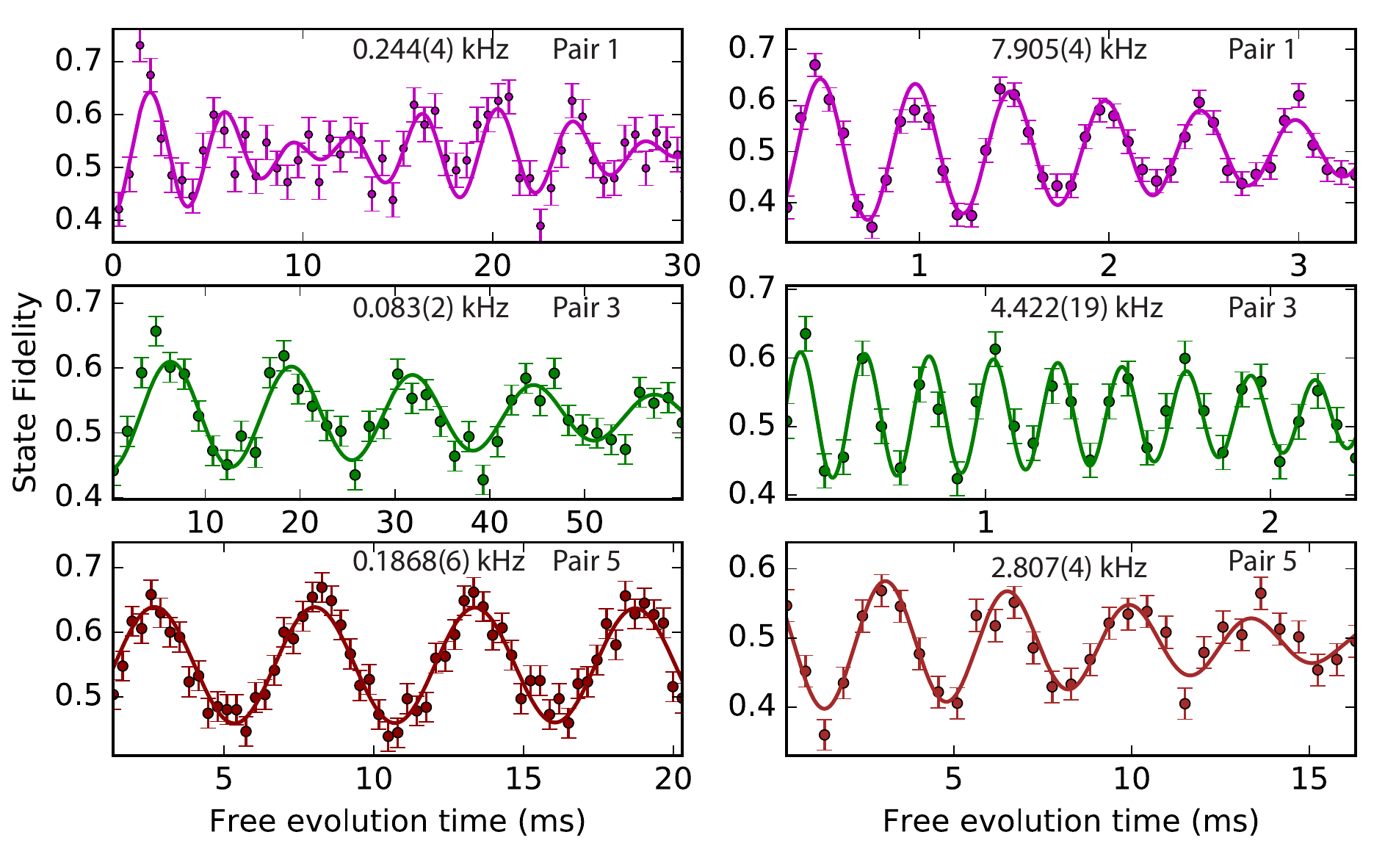}
\caption{\label{Figure 1}
\textbf{Direct spectroscopy of $^{13}C$-$^{13}C$ pairs.}  Ramsey spectroscopy for pairs 1, 3 and 5 and for electron state $m_s=0$ (left) and $m_s=-1$ (right) during the free evolution time. The measurement sequence is shown in Fig. 3a. These pairs are all of the type $Z >> X$. For the measurements with $m_s = -1$ an artificial detuning was applied. Pair 1 shows an additional beating (frequency of 22(2) Hz) indicating a small coupling to one (or more) additional spins.  Parameters and fit results are summarized in Supplementary Table \ref{Tab:Clusters}.}
\end{figure*}

\begin{figure*}[tb]
\centering
\includegraphics[scale=0.68]{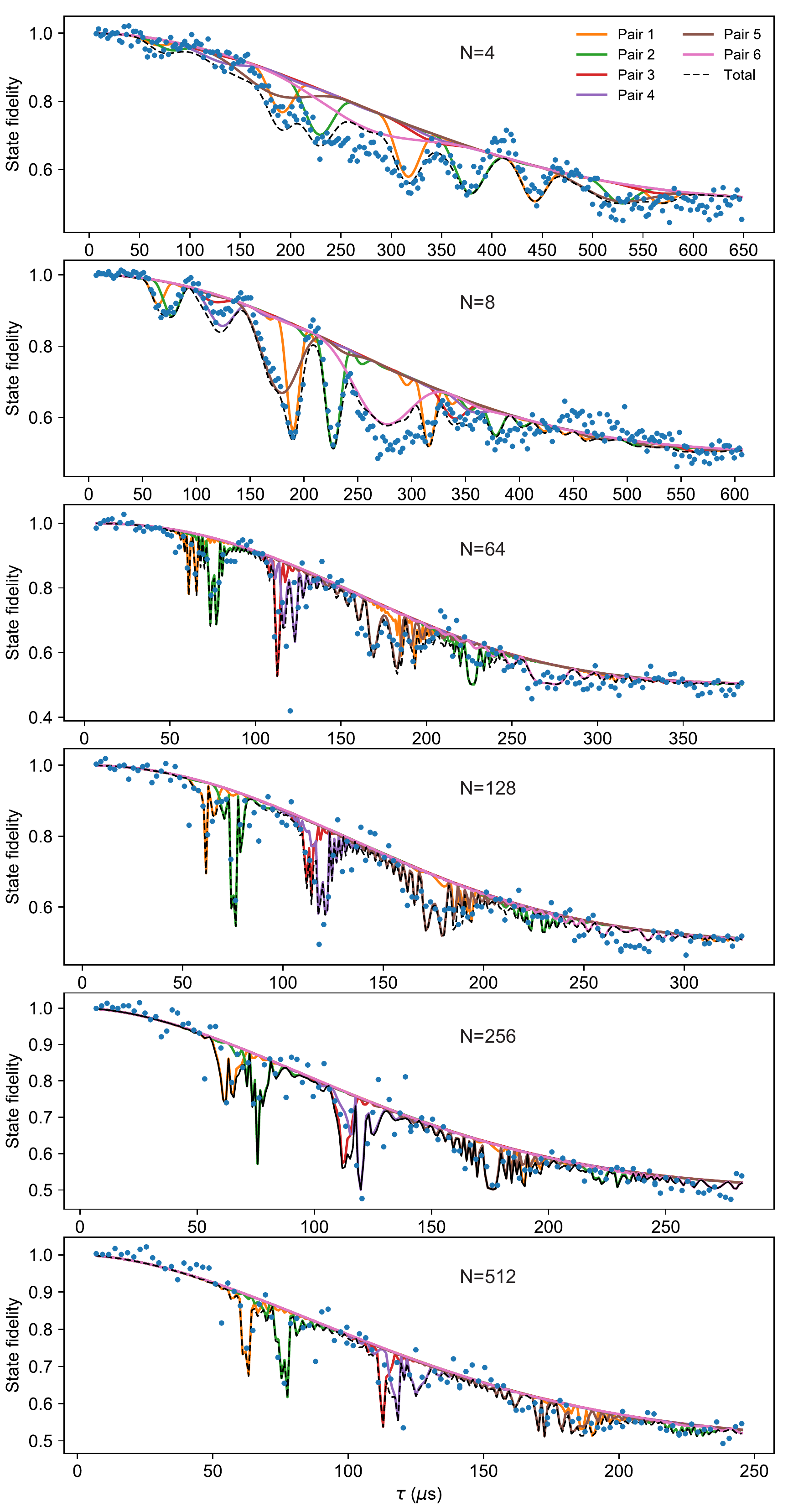}
\caption{\label{Figure 2}
\textbf{Comparison of the theory for the six $^{13}C-^{13}C$ pairs to the decoupling data.} Similar to the examples in Fig. 4 for N = 16 and 32, here we show extra examples for different N to confirm that the six identified $^{13}C-^{13}C$ pairs provide a good description of the dynamical decoupling data. $\tau = m \cdot \frac{2\pi}{\omega_L}$ to avoid coupling to individual $^{13}C$ spins.}
    \end{figure*}
 
\begin{figure*}[tb]
\centering
\includegraphics[scale=0.85]{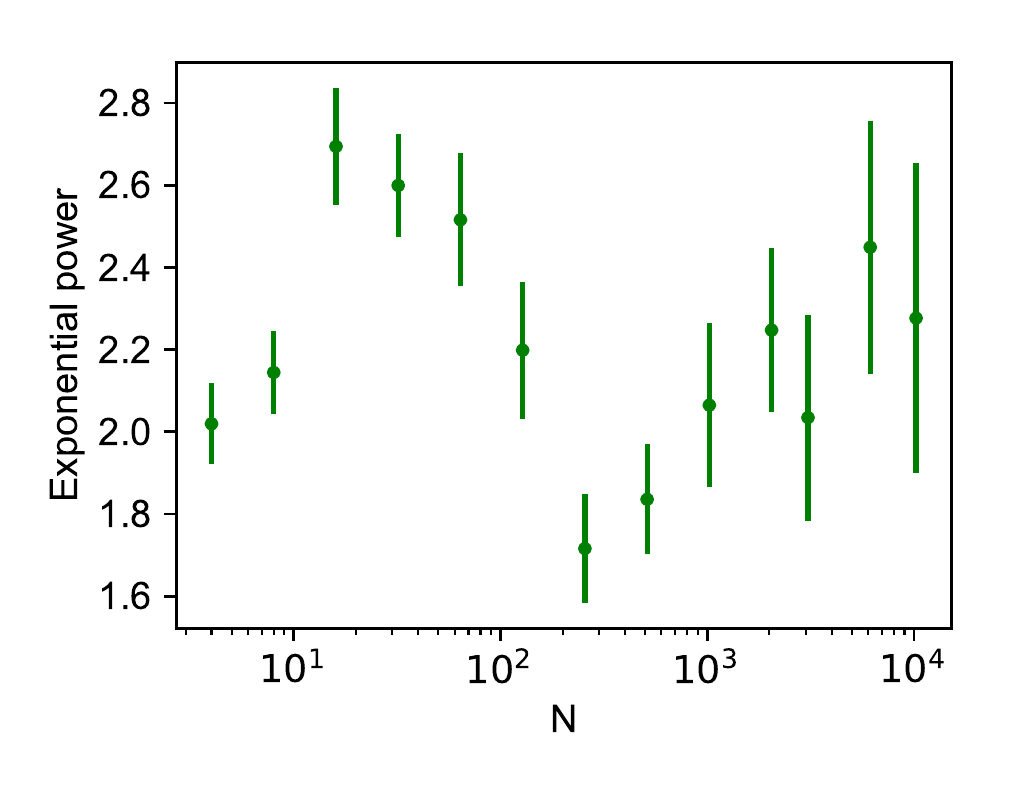}
\caption{\label{Figure_power}
\textbf{Extracted exponent of the coherence decay.} Fitted values of $n$ for the $e^{-(\tau/T)^n}$ envelop decay for the different numbers of pulses $N$ in Fig. 5a. The fact that the value is around 2 even for $N = 10^4$ pulses confirms that coherence times are not yet $T_1$ limited (expected $n = 1$ for $T_1$ limited case). } 
    \end{figure*}

\begin{figure*}[tb]
\centering
\includegraphics[scale=0.85]{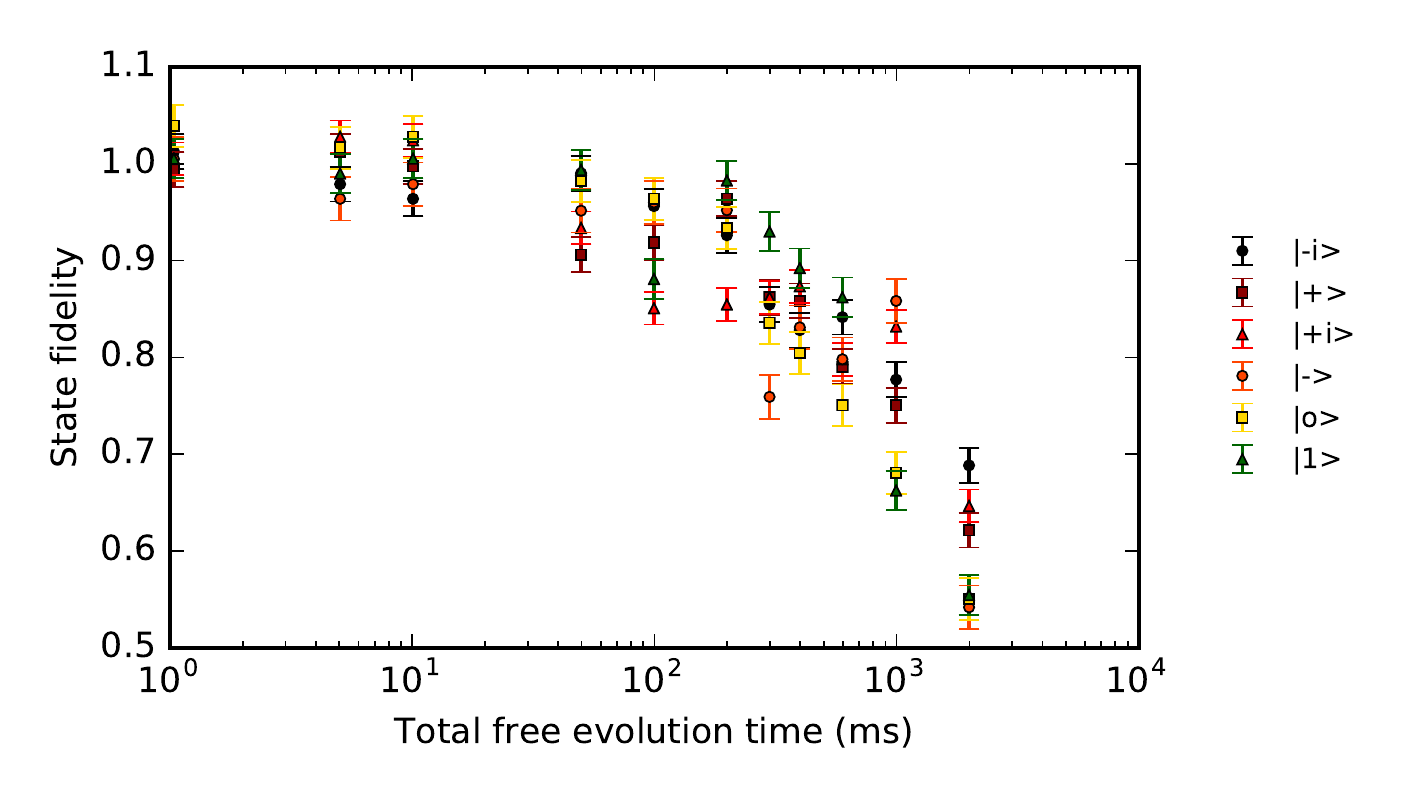}
\caption{\label{Figure 4}
\textbf{State fidelity for the six states used in Fig. 5c.} We prepare the six states: $|0\rangle = |m_s=0\rangle$, $|1\rangle = |m_s=-1\rangle$, $|\pm\rangle = (|0\rangle \pm|1\rangle)$, $|\pm i\rangle = (|0\rangle \pm i|1\rangle)$ and measure the fidelity of the final state with the ideal initial state. The curve in Fig. 5c in the main text is the average of these six state fidelities. The spin eigenstates $|0\rangle$ and $|1\rangle$ show a similar decay to the superposition states, indicating that the fidelities are likely limited by pulse errors.}
    \end{figure*}
\clearpage

\begin{table}[h!]
  \centering
  \begin{tabular}{c|c}
  	   
       $m_s=-1$ transition ($f_{-1}$)		& 1.746666(3) GHz 		\\
       $m_s=+1$ transition ($f_{+1}$)		& 4.008580(3) GHz 		\\ 	
       Zero field splitting ($\Delta$) 	& 2.877623 GHz 			\\ [1mm]
       \hline 												
       Magnetic field $B_z$ 	& 403.553 G   			\\ 
       Magnetic field stability & $3\cdot 10^{-3}$ G  	\\ 
       Magnetic field alignment & $ < 0.35 $  degrees 	\\ [1mm]
       \hline
       Electron $T_2^*$			& 4.9(2) $\mu$s			\\
	   Electron $T_2$ 			& 1.182(5) ms				\\
       Electron $T_1$ 			& $3.6(3) \cdot 10^3$ s 	\\	[1mm]
       \hline
       Frequency for pulse spacing $\omega_L/2\pi$ & $432.004$ kHz \\
       Period for pulse spacing $\tau_L$ & $2.3147\ \mu$s \\ 
       \hline
       Electron Rabi freq. & 14.31(3) MHz \\
       \hline
       NV strain 		& 4.0 GHz\\
   \end{tabular}
  \caption{\textbf{Experimental parameters.} $m_s =-1$ and $m_s = +1$ transitions are the obtained frequencies from electron spin resonance (ESR) measurements for the two spin transitions $0 \to -1$ and $0 \to +1$.  Assuming a well-aligned field with the NV axis, the zero field splitting (ZFS) is the average of the two frequencies. The magnetic field $B_z$ is the estimated field strength from $m_s = -1$ and $m_s = +1$ frequencies ($B_z = \frac{f_{+1} - f_{-1}} {2\gamma_e}, \gamma_e = 2.8024$ MHz/G). The magnetic field stability is the standard deviation of the magnetic field measured continuously (typical measurement time is 80 s) for 5 hours, during which the magnetic field is re-calibrated every 30 minutes, just as during the actual measurements. The magnetic field is aligned with the NV axis by sweeping the magnet position in the transversal directions and minimizing $\frac{f_{+1} + f_{-1}} {2}$. The maximum deviation from the minimum is estimated to be $10$ kHz. This implies a maximum perpendicular field $B_{\perp}$ of $2.5$ G, equivalent to a maximum misalignment angle of $0.35$ degrees. Electron $T_2^{*} $ is the free induction decay of the electron spin measured in a Ramsey interference experiment. Electron $T_2$ is the 1/e decay time of a spin echo measurement. Electron $T_1$ is the electron relaxation time shown in Fig. 1. $\omega_L/2\pi$ is the $^{13}C$ bare Larmor frequency estimated from the ESR measurements ($\omega_L = 2\pi\gamma_c \cdot B_z$, $\gamma_c = 1.0705$ kHz/G). $\tau_L$ is the estimated $^{13}C$ Larmor period ($\frac{2\pi}{\omega_L}$) as used for setting the half interpulse delay ($\tau= m \cdot \tau_L$) in the dynamical decoupling sequences. NV strain is the splitting of the excited states $E_x$ and $E_y$ due to strain perpendicular to the NV axis, measured by a resonant excitation spectroscopy at a temperature of 3.7 K.}
 \label{Tab:ExpParams}
\end{table}

\begin{table}[h!]
  \centering
  \begin{tabular}{c|c|c|c|c|c}
  			& $\omega_0 / 2\pi$ (kHz) & $\omega_1 / 2\pi$ (kHz) & $T_2^{*}$ (ms)&	$A_\parallel /2\pi$ (kHz) & $A_\perp/2\pi$(kHz) \\
            \hline
   C1	&   431.994(3) 	&	469.320(5)					& 10.2(4) & -36.4   &	25 	\\
   C2 	& 	431.874(3)	&	413.739(1)					& 12.5(5) &	 20.6 	& 	43	\\
   C3	&	431.891(2)						&   447.209(2)  				&  6.6(3) & -11.0   &	55	\\
   C4	&	431.947(3)						&   440.740(2) & 8.3 (6) &	 8.1	&   21	\\
   C5	&	431.934(3)	& 408.303(3)					& 20.8(7) &	 24.4 	&	26	\\
   C6	&	431.960(1)						&	480.607(4)  				& 4.0(2)  &	-48.7 	&	12	\\
   C7	&	431.95(1)						&	446.63(1)& 5.0(7)  &	 14.5 	&	11	\\
   \end{tabular}
  \caption{\textbf{Spectroscopy of isolated $^{13}C$.}  $A_\parallel$ and $A_\perp$ are the estimated hyperfine interaction components parallel and perpendicular to the applied magnetic field, obtained by matching the decoupling signal (e.g. Fig. 2a, b) to the theoretical coherence dips for the spins \cite{Taminiau_PRL2012_S}. Uncertainties are estimated to be of the order of the last digit. $\omega_0$ and $\omega_1$ are the measured nuclear precession frequencies for $m_s = 0$ and $m_s = -1$ ($m_s = +1$ for C4 and C7). $T_2^{*}$ is the dephasing time for $m_s= -1$ ($m_s= +1$ for C4 and C7).  We use the variation in $\omega_0$ for these spins as an estimate of how much $\omega_0$ varies between spins. This variation can be explained by an effective g-tensor for $^{13}C$ nuclear spins, due to a slightly misaligned field (see Supplementary Note 3). We study how these variations can affect the dynamical decoupling signal in Supplementary Fig. \ref{Figure_delta_f}.
}\label{Tab:Spins}
\end{table}
\clearpage

\begin{table}[h!]
  \centering
  \begin{tabular}{c|c|c|c}
   $\vec{r}$  &  $r$  &  $\theta$ (degrees)  &  $X/ 2\pi$ (Hz) \\[2mm]
   \hline
   $\dfrac{a_0}{4}$[1,1,1] & $0.433 a_0$  & 0 & 2061.0 
  	\\[2mm]
    \hline
   $\dfrac{a_0}{4}$[1,-1,1]& $0.433 a_0$  & 70.5 & 687.0 
  	\\[2mm]
        \hline
    $\dfrac{a_0}{4}$[2,2,0] & $0.707 a_0$  & 35.3 & 236.7
  	\\[2mm]
    \hline
   $\dfrac{a_0}{4}$[1,1,3] & $0.829 a_0$  & 29.5 & 186.8
  	\\[2mm]
    \hline
    $\dfrac{a_0}{4}$[1,-3,1] & $0.829 a_0$  & 80.0 & 133.4
  	\\[2mm]
        \hline
   $\dfrac{a_0}{4}$[3,1,3] & $1.089 a_0$  & 22.0 & 102.1
  	\\[2mm]
    \hline
   $\dfrac{a_0}{4}$[2,2,4] & $1.225 a_0$  & 19.5 & 75.9
  	\\[2mm]
    \hline
   $\dfrac{a_0}{4}$[3,-1,-3] & $1.089 a_0$  & 82.4 & 61.3
  	\\
   \end{tabular}
  \caption{\textbf{Main $^{13}C-^{13}C$ pairs in the diamond lattice and their calculated coupling strengths.}  The coupling strength $X$ is given by equation (\ref{eq:coupling_strength}). $\vec{r}$ is the vector connecting the $^{13}C-^{13}C$ pair; $\theta$ is the angle between the axis pair and 
  the external magnetic field which is along [1,1,1]. The distance between the two carbons forming the pair is $r$. The diamond lattice constant is taken to be 
  $a_0 = 3.5668$ \AA\ at 3.7 K \cite{diamond_lattice_S}. This table shows pairs with coupling strength down to 61 Hz which covers the range of pairs that we detected in this work.}
 \label{Tab:coupling_strength}
\end{table}

\begin{table}[h!]
  \centering
  \begin{tabular}{c|c|c|c|c|c|c}
  			& $\tau$ ($\mu$s) & N & $\omega_0/2\pi$ (kHz) & $X_{theory}/ 2\pi$ (kHz) &$\omega_{-1}/2\pi$ (kHz) & $Z/ 2\pi$ (kHz)  \\
            \hline
   Pair 1	& 63  & 14 & 0.244(3)	 & 0.2367	& 7.894(9)		& 7.890(9)   \\ 
   Pair 2 	& 76  & 10 & 0.247(6)	 & 0.2367	& 6.587(7)		& 6.582(6)    \\ 
   Pair 3	& 111 &	26 & 0.083(2)    & 0.0759	& 4.42(2)	    & 4.42(2)     \\
   Pair 4	& 120 &	24 & 2.0827(7)   & 2.061    & 2.0843(2)	    & 0.230	      \\ 
   Pair 5	& 172 &	8 & 0.1868(6)   & 0.1868	& 2.807(4)		& 2.801(4)     \\
   Pair 6	& 277 &	8  & 0.1338(1)   & 0.1334	& 1.831(3)		& 1.826(3)     \\
   \end{tabular}
  \caption{\textbf{Parameters for the six $^{13}C-^{13}C$ pairs.}  $\tau$ is half of the interpulse delay and $N$ is the number of pulses in the decoupling sequence used to perform the conditional gates in the Ramsey measurement sequences shown in Fig. 3. $\omega_0$ and $\omega_{-1}$ are the measured pseudo-spin precession frequencies for $m_s = 0$ and $m_s = -1$ respectively. $\omega_0/2\pi$ is a direct measurement of the coupling strength $X$ and $X_{theory}$ is the closest theoretical dipolar coupling strength to this value. This can be used to determine the atomic structure of the pair as shown in Supplementary Table \ref{Tab:coupling_strength}. $Z$ is due to the hyperfine field gradient and is calculated from the measured $\omega_0$ and $\omega_{-1}$: $Z = \sqrt{\omega_{-1}^{2}-\omega_0^{2}}$. Note that for pair 4 we have $X >> Z$, so the resonance condition is mainly governed by the coupling strength $X$ (resonant $\tau \sim 120\ \mu$s). Therefore, it is likely that additional pairs with the same $X$ \textemdash\ but smaller $Z$ values \textemdash\ contribute to the observed signal at $120\ \mu$s. Here we match the measured dynamical decoupling data for different values of $N$ (see e.g. Fig. 4b) to the calculated signal for a single pair, and find that the results are accurately reproduced for $Z/2\pi = 0.230$ kHz.
}\label{Tab:Clusters}
\end{table}
\clearpage


\begin{thebibliography}{10}
\expandafter\ifx\csname url\endcsname\relax
  \def\url#1{\texttt{#1}}\fi
\expandafter\ifx\csname urlprefix\endcsname\relax\def\urlprefix{URL }\fi
\providecommand{\bibinfo}[2]{#2}
\providecommand{\eprint}[2][]{\url{#2}}

\bibitem{Pfaff_NatPhys2013}
\bibinfo{author}{Pfaff, W.} \emph{et~al.}
\newblock \bibinfo{title}{Demonstration of entanglement-by-measurement of
  solid-state qubits}.
\newblock \emph{\bibinfo{journal}{Nat. Phys.}}
  \href{https://doi.org/10.1038/nphys2444}{\textbf{\bibinfo{volume}{9}},
  \bibinfo{pages}{29--33}} (\bibinfo{year}{2013}).

\bibitem{Waldherr_Nature2014}
\bibinfo{author}{Waldherr, G.} \emph{et~al.}
\newblock \bibinfo{title}{Quantum error correction in a solid-state hybrid spin
  register}.
\newblock \emph{\bibinfo{journal}{Nature}} \textbf{\bibinfo{volume}{506}},
  \bibinfo{pages}{204--218} (\bibinfo{year}{2014}).

\bibitem{Cramer_NatureComm2016}
\bibinfo{author}{Cramer, J.} \emph{et~al.}
\newblock \bibinfo{title}{Repeated quantum error correction on a continuously
  encoded qubit by real-time feedback}.
\newblock \emph{\bibinfo{journal}{Nat. Commun.}} \textbf{\bibinfo{volume}{7}},
  \bibinfo{pages}{11526} (\bibinfo{year}{2016}).

\bibitem{Wolfowicz_NJP2016}
\bibinfo{author}{Wolfowicz, G.} \emph{et~al.}
\newblock \bibinfo{title}{$^{29}${Si} nuclear spins as a resource for donor
  spin qubits in silicon}.
\newblock \emph{\bibinfo{journal}{New J. Phys.}} \textbf{\bibinfo{volume}{18}},
  \bibinfo{pages}{023021} (\bibinfo{year}{2016}).

\bibitem{Muhonen_NatNano2014}
\bibinfo{author}{Muhonen, J.~T.} \emph{et~al.}
\newblock \bibinfo{title}{Storing quantum information for 30 seconds in a
  nanoelectronic device}.
\newblock \emph{\bibinfo{journal}{Nat. Nanotech.}}
  \textbf{\bibinfo{volume}{9}}, \bibinfo{pages}{986--991}
  (\bibinfo{year}{2014}).

\bibitem{dehollain2016bell}
\bibinfo{author}{Dehollain, J.~P.} \emph{et~al.}
\newblock \bibinfo{title}{Bell's inequality violation with spins in silicon}.
\newblock \emph{\bibinfo{journal}{Nat. Nanotech.}}
  \textbf{\bibinfo{volume}{11}}, \bibinfo{pages}{242--246}
  (\bibinfo{year}{2016}).

\bibitem{Zaier_NatComm2016}
\bibinfo{author}{Zaiser, S.} \emph{et~al.}
\newblock \bibinfo{title}{Enhancing quantum sensing sensitivity by a quantum
  memory}.
\newblock \emph{\bibinfo{journal}{Nat. Commun.}} \textbf{\bibinfo{volume}{7}},
  \bibinfo{pages}{12279} (\bibinfo{year}{2016}).

\bibitem{Pfender_arXiv2017}
\bibinfo{author}{Pfender, M.} \emph{et~al.}
\newblock \bibinfo{title}{Nonvolatile nuclear spin memory enables
  sensor-unlimited nanoscale spectroscopy of small spin clusters}.
\newblock \emph{\bibinfo{journal}{Nat. Commun.}} \textbf{\bibinfo{volume}{8}},
  \bibinfo{pages}{834} (\bibinfo{year}{2017}).

\bibitem{Rosskopf_NPJQI2017}
\bibinfo{author}{Rosskopf, T.}, \bibinfo{author}{Zopes, J.},
  \bibinfo{author}{Boss, J.~M.} \& \bibinfo{author}{Degen, C.~L.}
\newblock \bibinfo{title}{A quantum spectrum analyzer enhanced by a nuclear
  spin memory}.
\newblock \emph{\bibinfo{journal}{NPJ Quantum Information}}
  \textbf{\bibinfo{volume}{3}}, \bibinfo{pages}{33} (\bibinfo{year}{2017}).

\bibitem{Lovchinsky_Science2016}
\bibinfo{author}{Lovchinsky, I.} \emph{et~al.}
\newblock \bibinfo{title}{Nuclear magnetic resonance detection and spectroscopy
  of single proteins using quantum logic}.
\newblock \emph{\bibinfo{journal}{Science}} \textbf{\bibinfo{volume}{351}},
  \bibinfo{pages}{836--841} (\bibinfo{year}{2016}).

\bibitem{Unden_PRL2016}
\bibinfo{author}{Unden, T.} \emph{et~al.}
\newblock \bibinfo{title}{Quantum metrology enhanced by repetitive quantum
  error correction}.
\newblock \emph{\bibinfo{journal}{Phys. Rev. Lett.}}
  \textbf{\bibinfo{volume}{116}}, \bibinfo{pages}{230502}
  (\bibinfo{year}{2016}).

\bibitem{Kolkowitz_PRL2012}
\bibinfo{author}{Kolkowitz, S.}, \bibinfo{author}{Unterreithmeier, Q.~P.},
  \bibinfo{author}{Bennett, S.~D.} \& \bibinfo{author}{Lukin, M.~D.}
\newblock \bibinfo{title}{Sensing distant nuclear spins with a single electron
  spin}.
\newblock \emph{\bibinfo{journal}{Phys. Rev. Lett.}}
  \textbf{\bibinfo{volume}{109}}, \bibinfo{pages}{137601}
  (\bibinfo{year}{2012}).

\bibitem{Taminiau_PRL2012}
\bibinfo{author}{Taminiau, T.~H.} \emph{et~al.}
\newblock \bibinfo{title}{Detection and control of individual nuclear spins
  using a weakly coupled electron spin}.
\newblock \emph{\bibinfo{journal}{Phys. Rev. Lett.}}
  \href{https://doi.org/10.1103/PhysRevLett.109.137602}{\textbf{\bibinfo{volume}{109}},
  \bibinfo{pages}{137602}} (\bibinfo{year}{2012}).

\bibitem{Zhao_NatureNano2012}
\bibinfo{author}{Zhao, N.} \emph{et~al.}
\newblock \bibinfo{title}{Sensing single remote nuclear spins}.
\newblock \emph{\bibinfo{journal}{Nat. Nanotech.}}
  \textbf{\bibinfo{volume}{7}}, \bibinfo{pages}{657--662}
  (\bibinfo{year}{2012}).

\bibitem{Muller_NatComm2014}
\bibinfo{author}{M{\"u}ller, C.} \emph{et~al.}
\newblock \bibinfo{title}{Nuclear magnetic resonance spectroscopy with single
  spin sensitivity}.
\newblock \emph{\bibinfo{journal}{Nat. Commun.}} \textbf{\bibinfo{volume}{5}},
  \bibinfo{pages}{4703} (\bibinfo{year}{2014}).

\bibitem{Shi_NatPhys2014}
\bibinfo{author}{Shi, F.} \emph{et~al.}
\newblock \bibinfo{title}{Sensing and atomic-scale structure analysis of single
  nuclear-spin clusters in diamond}.
\newblock \emph{\bibinfo{journal}{Nat. Phys.}} \textbf{\bibinfo{volume}{10}},
  \bibinfo{pages}{21--25} (\bibinfo{year}{2014}).

\bibitem{Lee_NatNano2013}
\bibinfo{author}{Lee, S.-Y.} \emph{et~al.}
\newblock \bibinfo{title}{Readout and control of a single nuclear spin with a
  metastable electron spin ancilla}.
\newblock \emph{\bibinfo{journal}{Nat. Nanotech.}}
  \textbf{\bibinfo{volume}{8}}, \bibinfo{pages}{487--492}
  (\bibinfo{year}{2013}).

\bibitem{Kalb_NatureComm2016}
\bibinfo{author}{Kalb, N.} \emph{et~al.}
\newblock \bibinfo{title}{Experimental creation of quantum zeno subspaces by
  repeated multi-spin projections in diamond}.
\newblock \emph{\bibinfo{journal}{Nat. Commun.}} \textbf{\bibinfo{volume}{7}},
  \bibinfo{pages}{13111} (\bibinfo{year}{2016}).

\bibitem{Hensen_Nature2015}
\bibinfo{author}{Hensen, B.} \emph{et~al.}
\newblock \bibinfo{title}{Loophole-free bell inequality violation using
  electron spins separated by 1.3 kilometres}.
\newblock \emph{\bibinfo{journal}{Nature}} \textbf{\bibinfo{volume}{526}},
  \bibinfo{pages}{682--686} (\bibinfo{year}{2015}).

\bibitem{Sen_NatPhoton2016}
\bibinfo{author}{Yang, S.} \emph{et~al.}
\newblock \bibinfo{title}{High-fidelity transfer and storage of photon states
  in a single nuclear spin}.
\newblock \emph{\bibinfo{journal}{Nat. Photon.}} \textbf{\bibinfo{volume}{10}},
  \bibinfo{pages}{507--511} (\bibinfo{year}{2016}).

\bibitem{Reiserer_PRX2016}
\bibinfo{author}{Reiserer, A.} \emph{et~al.}
\newblock \bibinfo{title}{Robust quantum-network memory using
  decoherence-protected subspaces of nuclear spins}.
\newblock \emph{\bibinfo{journal}{Phys. Rev. X}}
  \href{https://doi.org/10.1103/PhysRevX.6.021040}{\textbf{\bibinfo{volume}{6}},
  \bibinfo{pages}{021040}} (\bibinfo{year}{2016}).

\bibitem{Kalb_Science2017}
\bibinfo{author}{Kalb, N.} \emph{et~al.}
\newblock \bibinfo{title}{Entanglement distillation between solid-state quantum
  network nodes}.
\newblock \emph{\bibinfo{journal}{Science}} \textbf{\bibinfo{volume}{356}},
  \bibinfo{pages}{928--932} (\bibinfo{year}{2017}).

\bibitem{Bar-Gill_NatComm2012}
\bibinfo{author}{Bar-Gill, N.}, \bibinfo{author}{Pham, L.~M.},
  \bibinfo{author}{Jarmola, A.}, \bibinfo{author}{Budker, D.} \&
  \bibinfo{author}{Walsworth, R.~L.}
\newblock \bibinfo{title}{Solid-state electronic spin coherence time
  approaching one second}.
\newblock \emph{\bibinfo{journal}{Nat. Commun.}} \textbf{\bibinfo{volume}{4}},
  \bibinfo{pages}{1743} (\bibinfo{year}{2013}).

\bibitem{Tyryshkin_NatMat2012}
\bibinfo{author}{Tyryshkin, A.~M.} \emph{et~al.}
\newblock \bibinfo{title}{Electron spin coherence exceeding seconds in
  high-purity silicon}.
\newblock \emph{\bibinfo{journal}{Nat. Mater.}} \textbf{\bibinfo{volume}{11}},
  \bibinfo{pages}{143--147} (\bibinfo{year}{2012}).

\bibitem{Wolfowicz_PRB2012}
\bibinfo{author}{Wolfowicz, G.} \emph{et~al.}
\newblock \bibinfo{title}{Decoherence mechanisms of $^{209}${Bi} donor electron
  spins in isotopically pure $^{28}${Si}}.
\newblock \emph{\bibinfo{journal}{Phys. Rev. B}}
  \href{https://doi.org/10.1103/PhysRevB.86.245301}{\textbf{\bibinfo{volume}{86}},
  \bibinfo{pages}{245301}} (\bibinfo{year}{2012}).

\bibitem{Wolfowicz_NatNano2013}
\bibinfo{author}{Wolfowicz, G.} \emph{et~al.}
\newblock \bibinfo{title}{Atomic clock transitions in silicon-based spin
  qubits}.
\newblock \emph{\bibinfo{journal}{Nat. Nanotech.}}
  \textbf{\bibinfo{volume}{8}}, \bibinfo{pages}{561--564}
  (\bibinfo{year}{2013}).

\bibitem{Takashi_PRL2008}
\bibinfo{author}{Takahashi, S.}, \bibinfo{author}{Hanson, R.},
  \bibinfo{author}{van Tol, J.}, \bibinfo{author}{Sherwin, M.~S.} \&
  \bibinfo{author}{Awschalom, D.~D.}
\newblock \bibinfo{title}{Quenching spin decoherence in diamond through spin
  bath polarization}.
\newblock \emph{\bibinfo{journal}{Phys. Rev. Lett.}}
  \href{https://doi.org/10.1103/PhysRevLett.101.047601}{\textbf{\bibinfo{volume}{101}},
  \bibinfo{pages}{047601}} (\bibinfo{year}{2008}).

\bibitem{Jarmola_PRL_2012}
\bibinfo{author}{Jarmola, A.}, \bibinfo{author}{Acosta, V.~M.},
  \bibinfo{author}{Jensen, K.}, \bibinfo{author}{Chemerisov, S.} \&
  \bibinfo{author}{Budker, D.}
\newblock \bibinfo{title}{Temperature- and magnetic-field-dependent
  longitudinal spin relaxation in nitrogen-vacancy ensembles in diamond}.
\newblock \emph{\bibinfo{journal}{Phys. Rev. Lett.}}
  \textbf{\bibinfo{volume}{108}}, \bibinfo{pages}{197601}
  (\bibinfo{year}{2012}).

\bibitem{Astner_arxiv2017}
\bibinfo{author}{Astner, T.} \emph{et~al.}
\newblock \bibinfo{title}{Solid-state electron spin lifetime limited by
  phononic vacuum modes}.
\newblock \emph{\bibinfo{journal}{ArXiv}} \bibinfo{pages}{1706.09798}
  (\bibinfo{year}{2017}).

\bibitem{Norambuena_arxiv2017}
\bibinfo{author}{Norambuena, A.} \emph{et~al.}
\newblock \bibinfo{title}{Spin-lattice relaxation of individual solid-state
  spins}.
\newblock \emph{\bibinfo{journal}{ArXiv}} \bibinfo{pages}{1711.10280}
  (\bibinfo{year}{2017}).

\bibitem{Zhao_NatureNat2011}
\bibinfo{author}{Zhao, N.}, \bibinfo{author}{Hu, J.-L.}, \bibinfo{author}{Ho,
  S.-W.}, \bibinfo{author}{Wan, J. T.~K.} \& \bibinfo{author}{Liu, R.~B.}
\newblock \bibinfo{title}{Atomic-scale magnetometry of distant nuclear spin
  clusters via nitrogen-vacancy spin in diamond}.
\newblock \emph{\bibinfo{journal}{Nat. Nanotech.}}
  \textbf{\bibinfo{volume}{6}}, \bibinfo{pages}{242--246}
  (\bibinfo{year}{2011}).

\bibitem{Childress_Science2006}
\bibinfo{author}{Childress, L.} \emph{et~al.}
\newblock \bibinfo{title}{Coherent dynamics of coupled electron and nuclear
  spin qubits in diamond}.
\newblock \emph{\bibinfo{journal}{Science}}
  \href{https://doi.org/10.1126/science.1131871}{\textbf{\bibinfo{volume}{314}},
  \bibinfo{pages}{281--285}} (\bibinfo{year}{2006}).

\bibitem{Wang_NatComm2017}
\bibinfo{author}{Wang, Z.-Y.}, \bibinfo{author}{Casanova, J.} \&
  \bibinfo{author}{Plenio, M.~B.}
\newblock \bibinfo{title}{Delayed entanglement echo for individual control of a
  large number of nuclear spins}.
\newblock \emph{\bibinfo{journal}{Nat. Commun.}} \textbf{\bibinfo{volume}{8}},
  \bibinfo{pages}{14660} (\bibinfo{year}{2017}).

\bibitem{Sar_Nature2012}
\bibinfo{author}{van~der Sar, T.} \emph{et~al.}
\newblock \bibinfo{title}{Decoherence-protected quantum gates for a hybrid
  solid-state spin register}.
\newblock \emph{\bibinfo{journal}{Nature}} \textbf{\bibinfo{volume}{484}},
  \bibinfo{pages}{82--86} (\bibinfo{year}{2012}).

\bibitem{Taminiau_NatNano2014}
\bibinfo{author}{Taminiau, T.~H.}, \bibinfo{author}{Cramer, J.},
  \bibinfo{author}{van~der Sar, T.}, \bibinfo{author}{Dobrovitski, V.~V.} \&
  \bibinfo{author}{Hanson, R.}
\newblock \bibinfo{title}{Universal control and error correction in multi-qubit
  spin registers in diamond}.
\newblock \emph{\bibinfo{journal}{Nat. Nanotech.}}
  \textbf{\bibinfo{volume}{9}}, \bibinfo{pages}{171--176}
  (\bibinfo{year}{2014}).

\bibitem{PhysRevX.5.021009}
\bibinfo{author}{Loretz, M.} \emph{et~al.}
\newblock \bibinfo{title}{Spurious harmonic response of multipulse quantum
  sensing sequences}.
\newblock \emph{\bibinfo{journal}{Phys. Rev. X}}
  \href{https://doi.org/10.1103/PhysRevX.5.021009}{\textbf{\bibinfo{volume}{5}},
  \bibinfo{pages}{021009}} (\bibinfo{year}{2015}).

\bibitem{PhysRevApplied.7.054009}
\bibinfo{author}{Lang, J.~E.}, \bibinfo{author}{Casanova, J.},
  \bibinfo{author}{Wang, Z.-Y.}, \bibinfo{author}{Plenio, M.~B.} \&
  \bibinfo{author}{Monteiro, T.~S.}
\newblock \bibinfo{title}{Enhanced resolution in nanoscale {NMR} via quantum
  sensing with pulses of finite duration}.
\newblock \emph{\bibinfo{journal}{Phys. Rev. Applied}}
  \href{https://doi.org/10.1103/PhysRevApplied.7.054009}{\textbf{\bibinfo{volume}{7}},
  \bibinfo{pages}{054009}} (\bibinfo{year}{2017}).

\bibitem{ajoy2016dc}
\bibinfo{author}{Ajoy, A.}, \bibinfo{author}{Liu, Y.} \&
  \bibinfo{author}{Cappellaro, P.}
\newblock \bibinfo{title}{Dc magnetometry at the $ t_2 $ limit}.
\newblock \emph{\bibinfo{journal}{ArXiv}} \bibinfo{pages}{1611.04691}
  (\bibinfo{year}{2016}).

\bibitem{Lange_science2010}
\bibinfo{author}{de~Lange, G.}, \bibinfo{author}{Wang, Z.~H.},
  \bibinfo{author}{Rist{\`e}, D.}, \bibinfo{author}{Dobrovitski, V.~V.} \&
  \bibinfo{author}{Hanson, R.}
\newblock \bibinfo{title}{Universal dynamical decoupling of a single
  solid-state spin from a spin bath}.
\newblock \emph{\bibinfo{journal}{Science}}
  \href{https://doi.org/10.1126/science.1192739}{\textbf{\bibinfo{volume}{330}},
  \bibinfo{pages}{60--63}} (\bibinfo{year}{2010}).

\bibitem{Ryan_PRL2010}
\bibinfo{author}{Ryan, C.~A.}, \bibinfo{author}{Hodges, J.~S.} \&
  \bibinfo{author}{Cory, D.~G.}
\newblock \bibinfo{title}{Robust decoupling techniques to extend quantum
  coherence in diamond}.
\newblock \emph{\bibinfo{journal}{Phys. Rev. Lett.}}
  \href{https://doi.org/10.1103/PhysRevLett.105.200402}{\textbf{\bibinfo{volume}{105}},
  \bibinfo{pages}{200402}} (\bibinfo{year}{2010}).

\bibitem{naydenov2011dynamical}
\bibinfo{author}{Naydenov, B.} \emph{et~al.}
\newblock \bibinfo{title}{Dynamical decoupling of a single-electron spin at
  room temperature}.
\newblock \emph{\bibinfo{journal}{Phys. Rev. B}} \textbf{\bibinfo{volume}{83}},
  \bibinfo{pages}{081201} (\bibinfo{year}{2011}).

\bibitem{Medford_PRL2012}
\bibinfo{author}{Medford, J.} \emph{et~al.}
\newblock \bibinfo{title}{Scaling of dynamical decoupling for spin qubits}.
\newblock \emph{\bibinfo{journal}{Phys. Rev. Lett.}}
  \href{https://doi.org/10.1103/PhysRevLett.108.086802}{\textbf{\bibinfo{volume}{108}},
  \bibinfo{pages}{086802}} (\bibinfo{year}{2012}).

\bibitem{Seo_NatComm2016}
\bibinfo{author}{Seo, H.} \emph{et~al.}
\newblock \bibinfo{title}{Quantum decoherence dynamics of divacancy spins in
  silicon carbide}.
\newblock \emph{\bibinfo{journal}{Nat. Commun.}} \textbf{\bibinfo{volume}{7}},
  \bibinfo{pages}{12935} (\bibinfo{year}{2016}).

\bibitem{Widmann_NatMat2014}
\bibinfo{author}{Widmann, M.} \emph{et~al.}
\newblock \bibinfo{title}{Coherent control of single spins in âsilicon
  carbide at room temperature}.
\newblock \emph{\bibinfo{journal}{Nat. Mater.}} \textbf{\bibinfo{volume}{14}},
  \bibinfo{pages}{164â168} (\bibinfo{year}{2014}).

\bibitem{Falk_PRL2015}
\bibinfo{author}{Falk, A.~L.} \emph{et~al.}
\newblock \bibinfo{title}{Optical polarization of nuclear spins in silicon
  carbide}.
\newblock \emph{\bibinfo{journal}{Phys. Rev. Lett.}}
  \href{https://doi.org/10.1103/PhysRevLett.114.247603}{\textbf{\bibinfo{volume}{114}},
  \bibinfo{pages}{247603}} (\bibinfo{year}{2015}).

\bibitem{Yang_PRB2014}
\bibinfo{author}{Yang, L.-P.} \emph{et~al.}
\newblock \bibinfo{title}{Electron spin decoherence in silicon carbide nuclear
  spin bath}.
\newblock \emph{\bibinfo{journal}{Phys. Rev. B}}
  \href{https://doi.org/10.1103/PhysRevB.90.241203}{\textbf{\bibinfo{volume}{90}},
  \bibinfo{pages}{241203}} (\bibinfo{year}{2014}).

\bibitem{Rogers_PRL2014}
\bibinfo{author}{Rogers, L.~J.} \emph{et~al.}
\newblock \bibinfo{title}{All-optical initialization, readout, and coherent
  preparation of single silicon-vacancy spins in diamond}.
\newblock \emph{\bibinfo{journal}{Phys. Rev. Lett.}}
  \href{https://doi.org/10.1103/PhysRevLett.113.263602}{\textbf{\bibinfo{volume}{113}},
  \bibinfo{pages}{263602}} (\bibinfo{year}{2014}).

\bibitem{Sukachev_Arxiv2017}
\bibinfo{author}{Sukachev, D.~D.} \emph{et~al.}
\newblock \bibinfo{title}{The silicon-vacancy spin qubit in diamond: quantum
  memory exceeding ten milliseconds and single-shot state readout}.
\newblock \emph{\bibinfo{journal}{ArXiv}} \bibinfo{pages}{1708.08852}
  (\bibinfo{year}{2017}).

\bibitem{Becker_Arxiv2017}
\bibinfo{author}{Becker, J.~N.} \emph{et~al.}
\newblock \bibinfo{title}{All-optical control of the silicon-vacancy spin in
  diamond at millikelvin temperatures}.
\newblock \emph{\bibinfo{journal}{ArXiv}} \bibinfo{pages}{:1708.08263}
  (\bibinfo{year}{2017}).

\bibitem{Rose_Arxiv2017}
\bibinfo{author}{Rose, B.~C.} \emph{et~al.}
\newblock \bibinfo{title}{Observation of an environmentally insensitive solid
  state spin defect in diamond}.
\newblock \emph{\bibinfo{journal}{ArXiv}} \bibinfo{pages}{:1706.01555}
  (\bibinfo{year}{2017}).

\bibitem{Siyushev_PRB2017}
\bibinfo{author}{Siyushev, P.} \emph{et~al.}
\newblock \bibinfo{title}{Optical and microwave control of germanium-vacancy
  center spins in diamond}.
\newblock \emph{\bibinfo{journal}{Phys. Rev. B}}
  \href{https://doi.org/10.1103/PhysRevB.96.081201}{\textbf{\bibinfo{volume}{96}},
  \bibinfo{pages}{081201}} (\bibinfo{year}{2017}).

\bibitem{Pla_PRL2014}
\bibinfo{author}{Pla, J.~J.} \emph{et~al.}
\newblock \bibinfo{title}{Coherent control of a single $^{29}\mathrm{Si}$
  nuclear spin qubit}.
\newblock \emph{\bibinfo{journal}{Phys. Rev. Lett.}}
  \href{https://doi.org/10.1103/PhysRevLett.113.246801}{\textbf{\bibinfo{volume}{113}},
  \bibinfo{pages}{246801}} (\bibinfo{year}{2014}).

\bibitem{Iwasaki_arXiv2017}
\bibinfo{author}{Iwasaki, T.} \emph{et~al.}
\newblock \bibinfo{title}{Tin-vacancy quantum emitters in diamond}.
\newblock \emph{\bibinfo{journal}{ArXiv}} \bibinfo{pages}{1708.03576}
  (\bibinfo{year}{2013}).

\bibitem{kucsko2013nanometer}
\bibinfo{author}{Kucsko, G.} \emph{et~al.}
\newblock \bibinfo{title}{Nanometer scale thermometry in a living cell}.
\newblock \emph{\bibinfo{journal}{Nature}} \textbf{\bibinfo{volume}{500}},
  \bibinfo{pages}{54} (\bibinfo{year}{2013}).

\bibitem{shi2015single}
\bibinfo{author}{Shi, F.} \emph{et~al.}
\newblock \bibinfo{title}{Single-protein spin resonance spectroscopy under
  ambient conditions}.
\newblock \emph{\bibinfo{journal}{Science}} \textbf{\bibinfo{volume}{347}},
  \bibinfo{pages}{1135--1138} (\bibinfo{year}{2015}).

\bibitem{tetienne2014nanoscale}
\bibinfo{author}{Tetienne, J.-P.} \emph{et~al.}
\newblock \bibinfo{title}{Nanoscale imaging and control of domain-wall hopping
  with a nitrogen-vacancy center microscope}.
\newblock \emph{\bibinfo{journal}{Science}} \textbf{\bibinfo{volume}{344}},
  \bibinfo{pages}{1366} (\bibinfo{year}{2014}).

\bibitem{Humphreys2017}
\bibinfo{author}{Humphreys, P.~C.} \emph{et~al.}
\newblock \bibinfo{title}{Deterministic delivery of remote entanglement on a
  quantum network}.
\newblock \emph{\bibinfo{journal}{Arxiv}} \bibinfo{pages}{1712.07567}
  (\bibinfo{year}{2017}).

\bibitem{Lidar_PRL1998}
\bibinfo{author}{Lidar, D.~A.}, \bibinfo{author}{Chuang, I.~L.} \&
  \bibinfo{author}{Whaley, K.~B.}
\newblock \bibinfo{title}{Decoherence-free subspaces for quantum computation}.
\newblock \emph{\bibinfo{journal}{Phys. Rev. Lett.}}
  \href{https://doi.org/10.1103/PhysRevLett.81.2594}{\textbf{\bibinfo{volume}{81}},
  \bibinfo{pages}{2594--2597}} (\bibinfo{year}{1998}).

\bibitem{Lieven}
\bibinfo{author}{Vandersypen, L. M.~K.} \& \bibinfo{author}{Chuang, I.~L.}
\newblock \bibinfo{title}{{NMR} techniques for quantum control and
  computation}.
\newblock \emph{\bibinfo{journal}{Rev. Mod. Phys.}}
  \href{https://doi.org/10.1103/RevModPhys.76.1037}{\textbf{\bibinfo{volume}{76}},
  \bibinfo{pages}{1037--1069}} (\bibinfo{year}{2005}).

\bibitem{warren1984effects}
\bibinfo{author}{Warren, W.~S.}
\newblock \bibinfo{title}{Effects of arbitrary laser or {NMR} pulse shapes on
  population inversion and coherence}.
\newblock \emph{\bibinfo{journal}{J. Chem. Phys.}}
  \textbf{\bibinfo{volume}{81}}, \bibinfo{pages}{5437--5448}
  (\bibinfo{year}{1984}).

\bibitem{XY8}
\bibinfo{author}{Gullion, T.}, \bibinfo{author}{Baker, D.~B.} \&
  \bibinfo{author}{Conradi, M.~S.}
\newblock \bibinfo{title}{New, compensated {Carr-Purcell} sequences}.
\newblock \emph{\bibinfo{journal}{Journal of Magnetic Resonance (1969)}}
  \href{https://doi.org/https://doi.org/10.1016/0022-2364(90)90331-3}{\textbf{\bibinfo{volume}{89}},
  \bibinfo{pages}{479 -- 484}} (\bibinfo{year}{1990}).

\bibitem{Robledo_Nature2011}
\bibinfo{author}{Robledo, L.} \emph{et~al.}
\newblock \bibinfo{title}{High-fidelity projective read-out of a solid-state
  spin quantum register}.
\newblock \emph{\bibinfo{journal}{Nature}}
  \href{https://doi.org/10.1038/nature10401}{\textbf{\bibinfo{volume}{477}},
  \bibinfo{pages}{574--578}} (\bibinfo{year}{2011}).

\bibitem{Hadden_APL2010}
\bibinfo{author}{Hadden, J.~P.} \emph{et~al.}
\newblock \bibinfo{title}{Strongly enhanced photon collection from diamond
  defect centers under microfabricated integrated solid immersion lenses}.
\newblock \emph{\bibinfo{journal}{App. Phys. Lett.}}
  \href{https://doi.org/10.1063/1.3519847}{\textbf{\bibinfo{volume}{97}},
  \bibinfo{pages}{241901}} (\bibinfo{year}{2010}).

\bibitem{Pfaff_Science2014}
\bibinfo{author}{Pfaff, W.} \emph{et~al.}
\newblock \bibinfo{title}{Unconditional quantum teleportation between distant
  solid-state quantum bits}.
\newblock \emph{\bibinfo{journal}{Science}}
  \href{https://doi.org/10.1126/science.1253512}{\textbf{\bibinfo{volume}{345}},
  \bibinfo{pages}{532--535}} (\bibinfo{year}{2014}).

\bibitem{Yeung_APL2012}
\bibinfo{author}{Yeung, T.~K.}, \bibinfo{author}{Sage, D.~L.},
  \bibinfo{author}{Pham, L.~M.}, \bibinfo{author}{Stanwix, P.~L.} \&
  \bibinfo{author}{Walsworth, R.~L.}
\newblock \bibinfo{title}{Anti-reflection coating for nitrogen-vacancy optical
  measurements in diamond}.
\newblock \emph{\bibinfo{journal}{App. Phys. Lett.}}
  \href{https://doi.org/10.1063/1.4730401}{\textbf{\bibinfo{volume}{100}},
  \bibinfo{pages}{251111}} (\bibinfo{year}{2012}).

\end{thebibliography}

\begin{thebibliography}{1}
\expandafter\ifx\csname url\endcsname\relax
  \def\url#1{\texttt{#1}}\fi
\expandafter\ifx\csname urlprefix\endcsname\relax\def\urlprefix{URL }\fi
\providecommand{\bibinfo}[2]{#2}
\providecommand{\eprint}[2][]{\url{#2}}

\bibitem{Zhao_NatureNano2012_S}
\bibinfo{author}{Zhao, N.} \emph{et~al.}
\newblock \bibinfo{title}{Sensing single remote nuclear spins}.
\newblock \emph{\bibinfo{journal}{Nat. Nanotech.}}
  \textbf{\bibinfo{volume}{7}}, \bibinfo{pages}{657--662}
  (\bibinfo{year}{2012}).

\bibitem{Shi_NatPhys2014_S}
\bibinfo{author}{Shi, F.} \emph{et~al.}
\newblock \bibinfo{title}{Sensing and atomic-scale structure analysis of single
  nuclear-spin clusters in diamond}.
\newblock \emph{\bibinfo{journal}{Nat. Phys.}} \textbf{\bibinfo{volume}{10}},
  \bibinfo{pages}{21--25} (\bibinfo{year}{2014}).

\bibitem{Zhao_NatureNat2011_S}
\bibinfo{author}{Zhao, N.}, \bibinfo{author}{Hu, J.-L.}, \bibinfo{author}{Ho,
  S.-W.}, \bibinfo{author}{Wan, J. T.~K.} \& \bibinfo{author}{Liu, R.~B.}
\newblock \bibinfo{title}{Atomic-scale magnetometry of distant nuclear spin
  clusters via nitrogen-vacancy spin in diamond}.
\newblock \emph{\bibinfo{journal}{Nat. Nanotech.}}
  \textbf{\bibinfo{volume}{6}}, \bibinfo{pages}{242--246}
  (\bibinfo{year}{2011}).

\bibitem{Childress_Science2006_S}
\bibinfo{author}{Childress, L.} \emph{et~al.}
\newblock \bibinfo{title}{Coherent dynamics of coupled electron and nuclear
  spin qubits in diamond}.
\newblock \emph{\bibinfo{journal}{Science}} \textbf{\bibinfo{volume}{314}},
  \bibinfo{pages}{281--285} (\bibinfo{year}{2006}).

\bibitem{Cramer_NatureComm2016_S}
\bibinfo{author}{Cramer, J.} \emph{et~al.}
\newblock \bibinfo{title}{Repeated quantum error correction on a continuously
  encoded qubit by real-time feedback}.
\newblock \emph{\bibinfo{journal}{Nat. Commun.}} \textbf{\bibinfo{volume}{7}},
  \bibinfo{pages}{11526} (\bibinfo{year}{2016}).

\bibitem{Taminiau_PRL2012_S}
\bibinfo{author}{Taminiau, T.~H.} \emph{et~al.}
\newblock \bibinfo{title}{Detection and control of individual nuclear spins
  using a weakly coupled electron spin}.
\newblock \emph{\bibinfo{journal}{Phys. Rev. Lett.}}
  \textbf{\bibinfo{volume}{109}}, \bibinfo{pages}{137602}
  (\bibinfo{year}{2012}).

\bibitem{diamond_lattice_S}
\bibinfo{author}{Stoupin, S.} \& \bibinfo{author}{Shvyd'ko, Y.~V.}
\newblock \bibinfo{title}{Thermal expansion of diamond at low temperatures}.
\newblock \emph{\bibinfo{journal}{Phys. Rev. Lett.}}
  \textbf{\bibinfo{volume}{104}}, \bibinfo{pages}{085901}
  (\bibinfo{year}{2010}).

\end{thebibliography}

\end{document}